\documentclass[aps,prx,reprint,amsmath,amssymb,superscriptaddress,longbibliography,flushbottom,floatfix]{revtex4-2}
\usepackage[T1]{fontenc}
\usepackage[english]{babel}
\addto\captionsenglish{}
\usepackage{graphicx}
\usepackage{amsthm}
\usepackage{txfonts} 
\usepackage{bm}

\usepackage[colorlinks=true,citecolor=blue,linkcolor=blue,urlcolor=blue]{hyperref}
\hypersetup{pdftitle={No-Go Theorems for Ordered Universal Entanglement Purification},pdfauthor={C. L. Liu}}
\makeatletter
\let\auto@bib@innerbib\@empty
\let\write@bibliographystyle\relax
\makeatother
\newtheorem{theorem}{Theorem}
\newtheorem{lemma}{Lemma}
\newtheorem{proposition}{Proposition}
\newtheorem{corollary}{Corollary}
\theoremstyle{definition}
\newtheorem{definition}{Definition}
\newcommand{\tr}{\operatorname{tr}}

\newcommand{\supp}{\operatorname{supp}}
\newcommand{\HS}{\mathrm{HS}}
\newcommand{\SEP}{\mathrm{SEP}}
\newcommand{\LOCC}{\mathrm{LOCC}}

\newcommand{\ket}[1]{|#1\rangle}
\newcommand{\bra}[1]{\langle#1|}
\newcommand{\E}{\mathcal E}
\newcommand{\Sset}{\mathcal S}
\begin{document}
\title{No-Go Theorems for Ordered Universal Entanglement Purification}
\author{C. L. Liu}
\email[Contact author: ]{clliu@sdu.edu.cn}
\affiliation{School of Information Science and Engineering, Shandong University, Qingdao 266237, China}

\begin{abstract}
Entanglement purification is a fundamental tool for obtaining high-fidelity entangled states used in various quantum technologies. It aims to increase the fidelity of noisy shared entangled states with a fixed maximally entangled state through local operations and classical communication (LOCC). In distributed quantum systems, such as networks employing quantum repeaters, limited memory, probabilistic entanglement generation, and unequal storage times can leave a finite collection of nonidentical pairs. Their states may remain incompletely characterized even when calibrated preparation and storage records establish their fidelity ordering. Recent work has investigated whether fidelity ordering permits nontrivial universal purification. A nontrivial universal $n$-to-1 protocol must, conditioned on success, never lower the highest input fidelity and strictly increase it for at least one allowed joint input. Although no-go results have been established for a restricted class of LOCC protocols, the corresponding question for general LOCC has remained unresolved. Here, we study ordered universal purification under separable operations. Specifically, for every fixed finite $n\ge2$, we prove that when independently prepared two-qubit pairs range over the full family of states above a fixed fidelity threshold $0\le F_*<1$, any separable success map that never lowers the highest input fidelity on the allowed ordered inputs must preserve that fidelity exactly whenever it succeeds. This excludes nontrivial universal LOCC purification even for arbitrarily high input fidelities, without imposing a positive lower bound on success probability. The impossibility persists with any fixed finite-dimensional pure catalyst returned exactly upon success, without requiring its return after failure. In contrast, an explicit 3-to-1 LOCC protocol achieves nontrivial universal purification on specific finite and infinite Bell-diagonal families containing the target Bell state and full-rank states, with all nonunit fidelities bounded away from one. These results establish the limits of fidelity ordering for universal purification and show that further restrictions on the allowed input family can enable universal improvement.
\end{abstract}
\maketitle

\section{Introduction}

Quantum entanglement is a fundamental resource for quantum information processing, including quantum communication, quantum cryptography, and distributed quantum computation~\cite{Bennett1993,Ekert1991,Cirac1999,Nickerson2014,Horodecki2009}. Reliable entanglement-based implementations of these tasks require entangled states of sufficiently high quality, often quantified by their fidelity with a maximally entangled state. However, during preparation, transmission, and storage, interactions between the quantum systems and their environment can degrade the quality of entanglement and lower this fidelity. To counteract this degradation, entanglement purification was introduced as a method for obtaining higher-fidelity entangled pairs from noisy pairs shared by distant parties. The parties process these pairs using local operations and classical communication (LOCC) and retain the resulting pairs when a specified success event occurs~\cite{Bennett1996,Deutsch1996,Duer2007,Chitambar2014}. By supplying higher-fidelity shared entanglement, purification supports these information-processing tasks and is a key component of quantum-repeater schemes for distributing entanglement over long distances~\cite{Briegel1998,Duan2001,Wehner2018}. Optical implementations based on local parity checks have been proposed~\cite{Pan2001}, and purification has been demonstrated with photons, trapped ions, and solid-state network nodes~\cite{Pan2003,Reichle2006,Kalb2017}.

Entanglement purification has been studied in both asymptotic and one-shot regimes. In the standard asymptotic setting, the objective is to determine the number of Bell pairs obtained per input copy in the limit of infinitely many identical copies and vanishing output error~\cite{Bennett1996QEC,Rains1999Rigorous}. For bipartite pure states, entanglement concentration achieves a rate equal to the von Neumann entropy of either reduced state~\cite{BennettConcentration1996}. For mixed states, recurrence protocols increase the fidelity through repeated probabilistic purification steps and, under ideal operations and suitable input conditions, can approach unit fidelity as the number of rounds increases~\cite{Bennett1996,Deutsch1996}. Hashing protocols establish achievable asymptotic purification rates, with coherent information providing a lower bound on the optimal rate under one-way LOCC~\cite{DevetakWinter2005}. Combining recurrence and hashing can improve achievable rates for Bell-diagonal resources~\cite{VollbrechtVerstraete2005}. Universal concentration protocols can also attain the asymptotic entropy rate without prior knowledge of the input pure state, provided that identical copies are available~\cite{MatsumotoHayashi2007}. Asymptotic purification theory has further been extended beyond independent and identically distributed sources to correlated sequences of bipartite states~\cite{BuscemiDatta2010}.

In the one-shot regime, the task is to characterize the entanglement extractable from a given finite resource at a prescribed output accuracy, without taking the number of input systems to infinity. The resource may comprise one or several shared pairs, and probabilistic protocols additionally require an assessment of the success probability. For pure states, deterministic and probabilistic local transformations are governed by the Schmidt coefficients~\cite{Nielsen1999,Vidal1999,LoPopescu2001}. Single-copy purification by local filtering provides a basic instance of this setting~\cite{Linden1998,Kent1998,Verstraete2001}. Finite-resource achievability and converse bounds have been developed for specified input resources~\cite{BuscemiDatta2010,FangDistillation2019}. Other developments include exact characterizations for pure states~\cite{Regula2019}, semidefinite formulations under positive-partial-transpose-preserving operations~\cite{Rains2001}, and one-shot rates under non-entangling maps~\cite{BrandaoDatta2011}. Practical optimization addresses the tradeoff between output fidelity and success probability~\cite{Rozpedek2018}, while circuit-design methods account for limited resources and noisy local operations~\cite{Fujii2009,Krastanov2019}. Standard one-shot optimizations allow the protocol to depend on the specified input state.

In practical implementations, only a finite number of shared entangled pairs may be available, and their states need not be identical or fully characterized. For example, probabilistic entanglement generation in quantum repeaters produces pairs at different times, so that unequal storage durations expose them to different amounts of memory decoherence~\cite{Santra2019,Wallnofer2024}. Repeater optimization and memory-cutoff strategies likewise account for the effect of waiting times on performance~\cite{Goodenough2021,LiCutoffs2021}. Knowledge of the shared states may also be limited to partial information, such as measured expectation values of selected observables~\cite{Brun2001}. These considerations motivate purification protocols that do not require a complete description of each shared state. Consider $n$ independently prepared pairs with states $\rho_1,\ldots,\rho_n$ belonging to a prescribed family $\Sset$, whose joint input state is $\rho_{\mathrm{in}}=\rho_1\otimes\cdots\otimes\rho_n$. A natural question is whether a single LOCC protocol, fixed throughout $\Sset$, can produce a two-qubit state $\rho_{\mathrm{out}}$, conditioned on a designated success outcome, such that $F(\rho_{\mathrm{out}})\ge\max_{1\le j\le n}F(\rho_j)$ for every allowed joint input with nonzero success probability. Here $F(\rho):=\tr(\Phi\rho)$, where $\Phi$ is a fixed, known maximally entangled two-qubit state. This requirement defines universal $n$-to-1 entanglement purification~\cite{Zang2025}. The protocol must succeed with positive probability on at least one allowed joint input. It is nontrivial if the inequality is strict for at least one such input. For any fixed finite $n\ge2$, no universal LOCC protocol exists over the family of all entangled two-qubit states. This impossibility already holds for the smaller family satisfying $F(\rho)>1/2$, when no fidelity ordering is supplied~\cite{Zang2025}. Related results clarify the roles of source assumptions and allowed operations. Purification of unknown mixed states has been developed under exchangeability assumptions, using measurement outcomes to update the state estimate~\cite{Brun2001}. Universal concentration has also been investigated for finitely many identical copies of an unknown two-qubit pure state, with optimal success probabilities established within specified classes of protocols based on concatenated two-qubit operations~\cite{Orthey2026}. Moreover, allowing operations that preserve positivity under partial transposition makes universal purification possible for the full two-qubit family with $F(\rho)>1/2$~\cite{Zang2025}.

Incomplete knowledge of the shared states does not preclude knowledge of their fidelity ordering. For example, in quantum repeaters, heralding signals record the generation times of entangled pairs. When pairs are prepared under the same conditions and undergo the same monotonic fidelity decay during storage, these records determine their fidelity ordering through their different storage durations. Thus, relative fidelities can be available without complete state characterization~\cite{Santra2019,Wallnofer2024,Zang2025}. These common preparation and storage conditions may also restrict the noise model. The ordered task considered here isolates the use of the fidelity ordering within a specified input family, without assuming additional common noise parameters. Following the ordered setting introduced in Ref.~\cite{Zang2025}, we label the independently prepared pairs according to $F(\rho_1)\ge\cdots\ge F(\rho_n)$ and ask whether a fixed LOCC protocol can achieve nontrivial universal purification under this restriction. A fixed protocol must produce a state satisfying $F(\rho_{\mathrm{out}})\ge F(\rho_1)$ for every allowed joint input with nonzero success probability, with strict inequality for at least one such input. Simply retaining the first pair satisfies the fidelity-nondecrease requirement. The physical question is whether processing the additional pairs can improve its fidelity while preserving this universal requirement.

A partial answer is available for bilocal Clifford entanglement purification (biCEP) protocols, which use conjugate local Clifford operations followed by local measurements and classical comparison of their outcomes~\cite{Jansen2022,Zang2025}. For complete families of entangled Bell-diagonal states, every universal biCEP protocol with ordered fidelity has successful output fidelity exactly equal to the highest input fidelity. Here completeness requires every member to be entangled, the family to contain $\Phi$, and each of the three other Bell states to have a nonzero coefficient in at least one member. This excludes nontrivial universal biCEP even for families with arbitrarily high fidelity thresholds~\cite{Zang2025}. The ordered problem for general LOCC is not settled by these biCEP results. A related open problem concerns universality without fidelity ordering on the full family of two-qubit states satisfying $F(\rho)>F_*$ for a higher threshold $1/2<F_*<1$~\cite{Zang2025}. The search for such protocols beyond biCEP, including under separable operations, is explicitly identified as a future direction in Chapter~5 of Ref.~\cite{ZangThesis2026}.

In this work, we establish a no-go theorem for ordered universal purification under general LOCC. For every fixed finite $n\ge2$ and every threshold $0\le F_*<1$, we show that any separable success map that is universal with ordered fidelity on the family of all two-qubit states satisfying $F(\rho)>F_*$ must satisfy $F(\rho_{\mathrm{out}})=F(\rho_1)$ whenever it succeeds. This includes the higher thresholds $1/2<F_*<1$ considered above. Since LOCC operations are separable, this excludes nontrivial universal LOCC purification even for arbitrarily high input fidelities, without imposing a positive lower bound on the success probability. The same analysis excludes every nonzero universal separable success map on these threshold families when no ordering is supplied. The proof establishes a general restriction on separable success maps by requiring fidelity nondecrease for a sequence of full-rank Bell-diagonal states approaching $\Phi$, while the remaining pairs are held in a fixed full-rank state. We further prove that the impossibility persists with any fixed finite-dimensional pure catalyst that is returned exactly upon success, without requiring its return after failure. Finally, we construct an explicit 3-to-1 LOCC protocol that achieves nontrivial universal purification with ordered fidelity on both finite and infinite complete Bell-diagonal families. These families contain $\Phi$, but the fidelities of their other members are bounded away from unity. The constructions demonstrate that completeness alone, although sufficient for the biCEP no-go theorem, does not exclude universal purification by general LOCC.

\section{Preliminaries and problem formulation}
\label{sec:setting}

We study an $n$-to-1 entanglement purification task in which Alice and Bob process $n$ independently prepared qubit pairs into one output pair, with the input pairs labeled in nonincreasing order of fidelity. All Hilbert spaces are finite dimensional. This section introduces the states and allowed operations, then defines the conditional fidelity requirement and the input-family assumptions used in the proofs.

\subsection{Bipartite states and Bell-state fidelity}

Alice holds qubits $A_1,\ldots,A_n$ and Bob holds qubits $B_1,\ldots,B_n$. Their joint local systems are denoted by $A:=A_1\cdots A_n$ and $B:=B_1\cdots B_n$. The output of purification is assigned to fixed qubits $\widehat A\widehat B$. If a protocol retains different input pairs for different classical outcomes, local swaps can transfer the retained pair to these output systems. A bipartite state $\rho_{AB}$ is a positive semidefinite operator of unit trace on $\mathcal H_A\otimes\mathcal H_B$. It is separable across $A:B$ if
\begin{equation}
 \rho_{AB}=\sum_\ell v_\ell\rho^A_\ell\otimes\rho^B_\ell,
 \qquad v_\ell\ge0,\quad\sum_\ell v_\ell=1,
 \label{eq:separablestate}
\end{equation}
where $\rho^A_\ell$ and $\rho^B_\ell$ are states on $\mathcal H_A$ and $\mathcal H_B$, respectively. A state without such a decomposition is entangled~\cite{Werner1989,Horodecki2009}. The support $\supp\rho_{AB}$ is the span of its eigenvectors with nonzero eigenvalues, and its rank is the dimension of this support. Full rank means $\rho_{AB}>0$ on the specified joint space $\mathcal H_A\otimes\mathcal H_B$. In particular, a full-rank state of one qubit pair is positive definite on its four-dimensional joint space. Full rank of both reduced states does not imply full rank of the joint state. For a nonzero bipartite pure vector, its Schmidt rank is the number of nonzero coefficients in its Schmidt decomposition, whether or not the vector is normalized~\cite{Horodecki2009,TerhalHorodecki2000}. A pure vector is a product vector exactly when this rank is one.

The desired output of purification is the Bell state
\begin{equation}
 \ket\phi:=\frac{\ket{00}+\ket{11}}{\sqrt2},
 \qquad \Phi:=\ket\phi\bra\phi.
 \label{eq:target}
\end{equation}
Here $\ket0,\ket1$ are fixed local computational basis vectors. We write $I_d$ for the identity operator on a $d$-dimensional space and $I_{AB}$ for the identity on $\mathcal H_A\otimes\mathcal H_B$. An unsubscripted $I$ acts on the space determined by its expression. Superscripts $T$, $*$, and $\dagger$ denote transpose, complex conjugation, and adjoint in the specified bases, respectively. We use the same basis convention to identify $\Phi$ on each input pair and on $\widehat A\widehat B$. The fidelity of a two-qubit state $\rho$ with this fixed pure state is
\begin{equation}
 F(\rho):=\tr(\Phi\rho)=\bra\phi\rho\ket\phi.
 \label{eq:fidelity}
\end{equation}
This is the squared-fidelity convention~\cite{Uhlmann1976,Jozsa1994}. In entanglement theory, an overlap with a specified Bell state is often called a singlet fraction after a fixed local change of Bell basis~\cite{Horodecki2009,Horodecki1999Teleportation}. To avoid ambiguity with optimized definitions, we use only the explicitly defined $F(\rho)$. The fully entangled fraction is defined by $F_{\mathrm{ent}}(\rho):=\max_{\ket\psi\,\mathrm{maximally\ entangled}}\bra\psi\rho\ket\psi$~\cite{Bennett1996QEC,Horodecki1999Teleportation}, whereas $F(\rho)$ uses the fixed target $\Phi$. Every separable two-qubit state obeys $F(\rho)\le1/2$~\cite{Horodecki1999Teleportation}. Thus $F(\rho)>1/2$ is sufficient, but not necessary, for entanglement.

Let $P_x,P_y,P_z$ be the Pauli matrices in the chosen computational basis, and set $\ket{\phi_\alpha}:=(P_\alpha\otimes I_2)\ket\phi$ and $\Phi_\alpha:=\ket{\phi_\alpha}\bra{\phi_\alpha}$ for $\alpha\in\{x,y,z\}$. Together with $\ket\phi$, the three vectors $\ket{\phi_\alpha}$ form an orthonormal Bell basis. A Bell-diagonal state has the form
\begin{equation}
 \rho=F(\rho)\Phi+\sum_{\alpha=x,y,z}q_\alpha\Phi_\alpha,
 \quad q_\alpha\ge0,\quad\sum_\alpha q_\alpha=1-F(\rho).
 \label{eq:bds}
\end{equation}
The four numbers $F(\rho),q_x,q_y,q_z$ are called its Bell coefficients and are also its four eigenvalues. The state is full rank if and only if all four are positive. It is entangled if and only if its largest Bell coefficient exceeds $1/2$~\cite{Horodecki2009}. This follows from the positive partial transpose criterion: partial transposition on Bob's qubit gives eigenvalues $1/2-F(\rho)$ and $1/2-q_\alpha$, and a two-qubit state is separable exactly when its partial transpose is positive semidefinite~\cite{Peres1996,Horodecki1996Separability}.

An important example is the isotropic family~\cite{HorodeckiReduction1999},
\begin{equation}
 \omega_f:=f\Phi+\frac{1-f}{3}(I_4-\Phi),
 \qquad 0\le f\le1.
 \label{eq:isotropic}
\end{equation}
Its Bell coefficients are $f,(1-f)/3,(1-f)/3,(1-f)/3$, so $F(\omega_f)=f$, it is full rank for $0<f<1$, and it is entangled precisely when $f>1/2$. It is invariant under $U\otimes U^*$ for every qubit unitary $U$.

\subsection{Separable operations, LOCC, and biCEP}

A quantum operation is a completely positive (CP), trace-nonincreasing linear map. Complete positivity means that the map remains positive when tensored with the identity map on any finite reference system~\cite{Kraus1971,Choi1975,Watrous2018}. A separable operation from $AB$ to $\widehat A\widehat B$ admits a product Kraus representation~\cite{Rains1997,Horodecki2009,Chitambar2014},
\begin{equation}
 \E(\rho)=\sum_{j=1}^{N}E_j\rho E_j^\dagger,
 \qquad E_j=A_j\otimes B_j,
 \label{eq:sep}
\end{equation}
where $A_j:\mathcal H_A\to\mathcal H_{\widehat A}$ and $B_j:\mathcal H_B\to\mathcal H_{\widehat B}$. Trace nonincrease is equivalent to
\begin{equation}
 \sum_{j=1}^{N}E_j^\dagger E_j\le I_{AB}.
 \label{eq:tracenonincreasing}
\end{equation}
The class of these operations is denoted by $\SEP$.

The finite product representation in Eq.~\eqref{eq:sep} is equivalent to separability of the Choi operator across the corresponding local output--reference systems. To state this equivalence and fix the reference-system convention, let $R_A$ and $R_B$ be copies of $A$ and $B$. Define the unnormalized vectors $\ket{\Omega_A}:=\sum_a\ket a_A\ket a_{R_A}$ and $\ket{\Omega_B}:=\sum_b\ket b_B\ket b_{R_B}$ in fixed orthonormal bases. The Choi operator is the operator representing $\E$ on these local input--reference pairs~\cite{Jamiolkowski1972,Choi1975,Watrous2018}:
\begin{equation}
 J_{\E}:=(\E\otimes\operatorname{id}_{R_AR_B})
 \bigl(\ket{\Omega_A}\bra{\Omega_A}\otimes
       \ket{\Omega_B}\bra{\Omega_B}\bigr).
 \label{eq:choi}
\end{equation}
Here $\operatorname{id}_{R_AR_B}$ is the identity map on the reference systems, distinct from their identity operator $I_{R_AR_B}$. Tensor factors are regrouped so that $\E$ acts on $AB$. The product Kraus representation gives, across $\widehat A R_A:\widehat B R_B$,
\begin{equation}
 J_{\E}=\sum_j\ket{A_j}\!\rangle\langle\!\bra{A_j}
              \otimes\ket{B_j}\!\rangle\langle\!\bra{B_j},
 \label{eq:choiproduct}
\end{equation}
where $\ket{A_j}\!\rangle:=(A_j\otimes I_{R_A})\ket{\Omega_A}$, and similarly for $B_j$. Conversely, suppose that $J_{\E}$ is separable across $\widehat A R_A:\widehat B R_B$. If $J_{\E}\ne0$, normalize it by its trace and apply Carath\'eodory's theorem to obtain a finite convex decomposition into pure product states~\cite[Prop.~6.6 and Cor.~6.7]{Watrous2018}. Absorbing the coefficients into the local vectors and reversing the above identification gives a finite product Kraus representation of the form in Eq.~\eqref{eq:sep}. The zero Choi operator corresponds to the zero map. Thus the finite representation and the Choi characterization describe the same class of operations. The trace condition can also be expressed as
\begin{equation}
 \tr_{\widehat A\widehat B}J_{\E}
   =\left(\sum_j E_j^\dagger E_j\right)^T\le I_{R_AR_B}.
 \label{eq:choitrace}
\end{equation}

LOCC operations are implemented by local quantum operations in the two laboratories and classical communication between them~\cite{Horodecki2009,Chitambar2014}. At each stage, one party performs a local measurement or channel and may communicate its classical result. Subsequent local operations can then depend on the complete communicated history. Local ancillary systems, local discarding, and shared classical randomness are allowed, whereas transmitting quantum systems is not. An adaptive protocol is one in which a later operation depends on earlier measurement results. After refining each local operation into individual Kraus terms, every fully specified sequence of such terms has a product Kraus operator, because every quantum operation acts within one laboratory. Combining sequences that are declared successful therefore gives a separable success map. Limits of LOCC success maps are also separable in finite dimensions, since the cone of separable Choi operators is closed~\cite{Chitambar2014}. General separable channels and instruments need not be implementable by LOCC with the same outcome probabilities~\cite{Bennett1999Nonlocality}.

The more restricted biCEP class used in Ref.~\cite{Zang2025} consists of the following operations~\cite{Jansen2022}. Alice applies an $n$-qubit Clifford unitary $C$ to $A$, while Bob applies $C^*$ to $B$. A Clifford unitary maps the Pauli group to itself under conjugation~\cite{Dehaene2003}. Here the $n$-qubit Pauli group consists of tensor products of $I_2,P_x,P_y,P_z$, multiplied by phases in $\{1,-1,i,-i\}$. Conjugate local Clifford operations permute products of Bell states up to phases, providing a useful description of this purification class~\cite{DehaenePermutations2003,Jansen2022}. Both parties measure qubits $2,\ldots,n$ in their local computational bases and compare their classical bit strings. They retain pair $A_1B_1$ exactly when the two bit strings agree. Otherwise the attempt fails. Thus every such protocol is LOCC, and the classes satisfy
\begin{equation}
 \mathrm{biCEP}\subseteq\LOCC\subseteq\SEP.
 \label{eq:operationinclusions}
\end{equation}
The first inclusion concerns the full measurement and output rule, not just the choice of local gates. Section~\ref{sec:counterexample} will use an LOCC protocol in which a classical choice and measurements determine which pair is retained and which results constitute success. Those rules differ from the biCEP definition above even though the local circuit uses Clifford gates.

A quantum instrument describes the operation together with its classical outcome. It is a collection of CP, trace-nonincreasing maps $\{\mathcal I_r\}$ whose sum is trace preserving~\cite{DaviesLewis1970,Watrous2018,Chitambar2014}. If $\mathcal R_{\mathrm s}$ is the set of outcomes declared successful, their combined operation is $\E:=\sum_{r\in\mathcal R_{\mathrm s}}\mathcal I_r$. We call $\E$ the success map. For the no-go proofs we allow every separable, trace-nonincreasing success map, even without assuming that it has a separable completion at its original success probability. Catalytic assistance, involving an additional shared state, is defined separately in Sec.~\ref{sec:catalysis}.

\subsection{Ordered universal purification and input families}

Let $\Sset$ be a specified family of two-qubit states. For ordered $n$-to-1 purification, $\rho_i$ denotes the state of input pair $A_iB_i$, whereas the protocol acts on their joint input state
\begin{equation}
 \rho:=\rho_{\mathrm{in}}=\rho_1\otimes\cdots\otimes\rho_n,
 \qquad \rho_i\in\Sset,
 \label{eq:productinput}
\end{equation}
with the pair fidelities ordered as
\begin{equation}
 F(\rho_1)\ge F(\rho_2)\ge\cdots\ge F(\rho_n).
 \label{eq:ordered}
\end{equation}
Independence is an assumption of the task and gives the tensor-product form in Eq.~\eqref{eq:productinput}. Each $\rho_i$ can nevertheless be entangled across $A_i:B_i$. In this subsection, $\rho=\rho_{\mathrm{in}}$ denotes the joint input, whereas $\rho_i$ denotes the state of one input pair. Later sections specify the systems of every newly introduced state, and use $X$ or $X_t$ for joint inputs when convenient. A family $\Sset$ and the fidelity ordering are specified when the protocol is chosen, but the particular states $\rho_i$ and their numerical fidelities are not supplied to it. The protocol may depend on $\Sset$ and $n$, but must be the same for every allowed ordered choice from that family. The comparison for successful purification is with the fidelity of the first pair.

For a normalized joint input state $\rho$, the probability that the protocol succeeds is
\begin{equation}
 p_{\E}(\rho):=\tr\E(\rho).
 \label{eq:successprobability}
\end{equation}
After all outcome-dependent corrections have been included in the instrument, we combine the outcomes declared successful and disregard the finer classical record. When the probability of this combined event is positive, its conditional output state is
\begin{equation}
 \rho_{\mathrm{out}}(\rho):=\frac{\E(\rho)}{p_{\E}(\rho)}.
 \label{eq:conditioned}
\end{equation}
Its overlap with $\Phi$ before normalization, including the success probability, is
\begin{equation}
 h_{\E}(\rho):=\tr[\Phi\E(\rho)].
 \label{eq:successnumerator}
\end{equation}
The conditional output fidelity is consequently
\begin{equation}
 F_{\E}(\rho):=F\bigl(\rho_{\mathrm{out}}(\rho)\bigr)
             =\frac{h_{\E}(\rho)}{p_{\E}(\rho)}.
 \label{eq:conditionalfidelity}
\end{equation}
To compare this fidelity with a specified number $f\in[0,1]$, define the unnormalized fidelity difference
\begin{equation}
 \Delta_{\E}(\rho;f):=h_{\E}(\rho)-f\,p_{\E}(\rho).
 \label{eq:fidelitydifference}
\end{equation}
For positive success probability, $\Delta_{\E}(\rho;f)=p_{\E}(\rho)[F_{\E}(\rho)-f]$. In ordered purification we choose $f=F(\rho_1)$, so the sign of $\Delta_{\E}(\rho;F(\rho_1))$ determines whether the conditional output improves, preserves, or lowers the highest input fidelity. If $p_{\E}(\rho)=0$, no conditional output state is defined. Nevertheless, positivity implies $\E(\rho)=0$ and hence $\Delta_{\E}(\rho;f)=0$.

\begin{definition}
\label{def:universal}
A success map $\E$ is universal with ordered fidelity on $\Sset$ if it has positive success probability on at least one allowed input and every input in Eq.~\eqref{eq:productinput} satisfying Eq.~\eqref{eq:ordered} obeys
\begin{equation}
 \Delta_{\E}(\rho;F(\rho_1))\ge0.
 \label{eq:universality}
\end{equation}
It is nontrivial if this inequality is strict for at least one allowed input, and is fidelity preserving on the allowed inputs if equality holds for all of them.
\end{definition}

The inequality concerns the output conditioned on the combined success event, not a separate requirement on every finer measurement record. We allow zero success probability on individual inputs and impose no uniform positive lower bound. Requiring positive probability on at least one allowed input excludes operations that never succeed anywhere in the specified domain, even if their linear maps are nonzero. Discarding pairs $2,\ldots,n$ and retaining pair $1$ gives a fidelity-preserving universal protocol. Preserving the fidelity does not require preserving the state itself.

Following Ref.~\cite{Zang2025}, a Bell-diagonal family $\Sset$ is \emph{complete} if all its members are entangled, $\Phi\in\Sset$, and each vector $\ket{\phi_\alpha}$ belongs to the support of at least one member. For example, $\{\Phi,\omega_f\}$ is complete for every fixed $1/2<f<1$, because $\omega_f$ has all four Bell coefficients positive. Completeness does not imply convexity. A family is convex if every mixture of two of its members remains in the family. For a complete family, convexity therefore also requires all such mixtures to remain entangled.

The no-go proof requires allowed first-pair states approaching $\Phi$ while the other pairs can be held in a fixed full-rank state. The following condition ensures both requirements and will connect Theorem~\ref{thm:main} to ordered purification.
\begin{definition}
\label{def:segment}
An input family $\Sset$ satisfies the segment condition if there is a full-rank Bell-diagonal state $\sigma\in\Sset$ such that
\begin{equation}
 (1-\lambda)\Phi+\lambda\sigma\in\Sset
 \label{eq:segment}
\end{equation}
for every $\lambda\in[0,1]$.
\end{definition}
For instance, $\{\omega_f:f_0\le f\le1\}$ satisfies this condition with $\sigma=\omega_{f_0}$ for any $0<f_0<1$, by Eq.~\eqref{eq:isotropic}. Its members are all entangled when $f_0>1/2$. In contrast, for $1/2<f_0<1$, the complete family $\{\Phi,\omega_{f_0}\}$ contains no interior point of that segment. The segment condition does not require the rest of $\Sset$ to be Bell diagonal or convex. Section~\ref{sec:consequences} shows why completeness together with convexity is sufficient, and Sec.~\ref{sec:counterexample} shows why completeness alone is not.

Because the task concerns the state conditioned on success, multiplication of a success map by an input-independent positive constant preserves universality and nontriviality whenever the rescaled map remains trace nonincreasing. This scaling allows any finite separable success map to be implemented by LOCC at a possibly different success probability, without changing its conditional output. Thus it connects SEP and LOCC for the existence of a universal conditional transformation. For a linear operator $M$, let $\|M\|_\infty:=\max_{\|v\|=1}\|Mv\|$ be its operator norm. Omit zero terms in Eq.~\eqref{eq:sep} and define
\begin{equation}
 \gamma_j:=\|A_j\|_\infty^2\|B_j\|_\infty^2,
 \qquad \Gamma:=\sum_j\gamma_j>0.
 \label{eq:localscaling}
\end{equation}
Alice chooses $j$ with classical probability $\gamma_j/\Gamma$ and tells Bob the choice. They apply the local filters $A_j/\|A_j\|_\infty$ and $B_j/\|B_j\|_\infty$, respectively, declaring success only when both filters succeed. Each normalized filter $L$ satisfies $L^\dagger L\le I$ and can be completed by the failure operator $(I-L^\dagger L)^{1/2}$ to form a local measurement. The contribution for this classical choice is
\begin{equation}
 \frac{\gamma_j}{\Gamma}
 \frac{E_j\rho E_j^\dagger}{\gamma_j}
 =\frac1\Gamma E_j\rho E_j^\dagger.
 \label{eq:scaledsimulation}
\end{equation}
Summing over $j$ implements $\E/\Gamma$ by LOCC. The probability of choosing $j$ is independent of the input, whereas the probability that the filters succeed generally depends on it. Thus a finite separable success map and this LOCC realization have identical normalized output states conditioned on success and identical sets of inputs with positive success probability. The scaled map is trace nonincreasing because $\sum_jE_j^\dagger E_j\le\Gamma I$. This is the stochastic implementation of separable operations discussed in Ref.~\cite[Sec.~XI.B]{Horodecki2009}. However, implementing the rescaled success map does not imply that a separable channel is an LOCC channel or that the two classes achieve the same success probabilities. We use the product Kraus representation to prove the impossibility directly.

\section{Separable maps under fidelity nondecrease}
\label{sec:fidelitystructure}

To derive the no-go theorem in Sec.~\ref{sec:consequences}, we first determine how fidelity nondecrease along states approaching $\Phi$ constrains a separable success map. Consider a qubit pair $A_1B_1$ and finite-dimensional auxiliary systems $A'B'$, with joint Hilbert space $\mathcal H_{A'B'}:=\mathcal H_{A'}\otimes\mathcal H_{B'}$. Define $\eta:=\sum_{\alpha=x,y,z}r_\alpha\Phi_\alpha$, where $r_\alpha>0$ and $\sum_\alpha r_\alpha=1$, and $\rho_t:=t\Phi+(1-t)\eta$. For $0<t<1$, the state $\rho_t$ is full rank and has fidelity $F(\rho_t)=t$. As $t\to1$, the coefficients of $\Phi_x$, $\Phi_y$, and $\Phi_z$ vanish in the fixed ratios $r_x:r_y:r_z$. We hold a fixed auxiliary state $\tau>0$, with $\tr\tau=1$, while varying $t$. The state $\tau$ may be entangled across $A':B'$, and the auxiliary systems need not be qubits. Expressions such as $\rho_t\otimes\tau$ group the systems as $A_1B_1:A'B'$, whereas product Kraus operators group them by laboratory as $A_1A':B_1B'$. These expressions describe the same systems after the corresponding permutation of tensor factors.

\begin{theorem}
\label{thm:main}
Let $\E$ be a nonzero separable success map from $A_1A':B_1B'$ to a qubit pair, and let $\tau$ be a fixed full-rank state on $\mathcal H_{A'B'}$. Set $\rho_t:=t\Phi+(1-t)\sum_{\alpha=x,y,z}r_\alpha\Phi_\alpha$, with fixed $r_\alpha>0$ and $\sum_\alpha r_\alpha=1$. If there is a sequence $0<t_m<1$, with $t_m\to1$, such that
\begin{equation}
 \Delta_{\E}(\rho_{t_m}\otimes\tau;t_m)\ge0
 \quad\text{for every }m,
 \label{eq:boundarytest}
\end{equation}
then every nonzero operator in any finite product Kraus representation of $\E$ can be written as
\begin{equation}
 E_j=(U_j\otimes\bra{a_j})\otimes
       (U_j^*\otimes\bra{b_j}),
 \label{eq:rigidkraus}
\end{equation}
where $U_j$ is a qubit unitary and $\bra{a_j}$ and $\bra{b_j}$ are linear functionals on $A'$ and $B'$, respectively, with no normalization requirement.
\end{theorem}

We first establish two lemmas used in the proof.

\begin{lemma}
\label{lem:classify}
Let $E:=A\otimes B$, where $A$ and $B$ each map a local qubit and a finite-dimensional auxiliary system to an output qubit. Suppose that, for every $\ket\chi\in\mathcal H_{A'B'}$,
\begin{equation}
 E(\ket{\phi}\otimes\ket{\chi})
 \in\mathbb C\ket{\phi}.
 \label{eq:perfectconstraint}
\end{equation}
Then either $E$ annihilates this entire input subspace, or
\begin{equation}
 E=(M\otimes\bra a)\otimes(M^{-T}\otimes\bra b)
 \label{eq:reciprocal}
\end{equation}
for an invertible qubit matrix $M$. Here $M^{-T}:=(M^{-1})^T=(M^T)^{-1}$ denotes the inverse transpose.
\end{lemma}

\begin{proof}
Let $\{\ket\ell\}$ and $\{\ket m\}$ be orthonormal bases of $\mathcal H_{A'}$ and $\mathcal H_{B'}$, respectively. Define the $2\times2$ matrices
\begin{equation}
 A_\ell:=A(I_2\otimes\ket\ell),\qquad
 B_m:=B(I_2\otimes\ket m).
 \label{eq:localblocks}
\end{equation}
The product vectors $\ket\ell\otimes\ket m$ form an orthonormal basis of the joint auxiliary space $\mathcal H_{A'}\otimes\mathcal H_{B'}$. On the input $\ket\phi\otimes\ket\ell\otimes\ket m$, and after grouping the two output qubits together, $E$ acts as $(A_\ell\otimes B_m)\ket\phi$.

For a matrix $X$ expressed in fixed orthonormal bases, define vectorization by
\begin{equation}
 \operatorname{vec}(X):=\sum_{u,v}X_{uv}\ket u\otimes\ket v.
 \label{eq:vecdefinition}
\end{equation}
With this order of tensor factors, $\ket\phi=\operatorname{vec}(I_2)/\sqrt2$ and
\begin{equation}
 (R\otimes S)\operatorname{vec}(X)
 =\operatorname{vec}(RXS^T)
 \label{eq:vecidentity}
\end{equation}
for matrices of compatible dimensions~\cite{Watrous2018}. Indeed, the coefficient of $\ket u\otimes\ket v$ on either side is $\sum_{r,s}R_{ur}X_{rs}S_{vs}$. The transpose in this identity does not include complex conjugation.

It follows that $(A_\ell\otimes B_m)\ket\phi$ is proportional to $\ket\phi$ if and only if
\begin{equation}
 A_\ell B_m^T=c_{\ell m}I_2
 \quad\text{for all }\ell,m
 \label{eq:blocks}
\end{equation}
for some complex coefficients $c_{\ell m}$. If every coefficient vanishes, $E$ annihilates $\ket\phi\otimes\ket\ell\otimes\ket m$ for every $\ell,m$. Linearity then gives the same statement for any auxiliary vector $\ket\chi$, including vectors entangled across $A':B'$.

Otherwise, choose $\ell_0,m_0$ such that $c_{\ell_0m_0}\ne0$. The matrices $A_{\ell_0}$ and $B_{m_0}$ are invertible because their product in Eq.~\eqref{eq:blocks} is a nonzero multiple of the identity. Put $M:=A_{\ell_0}$. For every $m$, Eq.~\eqref{eq:blocks} gives
\begin{equation}
 MB_m^T=c_{\ell_0m}I_2,\qquad
 B_m=c_{\ell_0m}M^{-T}.
 \label{eq:Bblocks}
\end{equation}
In particular, $B_{m_0}^T=c_{\ell_0m_0}M^{-1}$. Substituting this into $A_\ell B_{m_0}^T=c_{\ell m_0}I_2$ yields
\begin{equation}
 A_\ell=\lambda_\ell M,\qquad
 \lambda_\ell:=\frac{c_{\ell m_0}}{c_{\ell_0m_0}}.
 \label{eq:Ablocks}
\end{equation}
Thus $\lambda_{\ell_0}=1$. The local operators can now be reconstructed from their matrices:
\begin{align}
 A&=\sum_\ell A_\ell(I_2\otimes\bra\ell)
   =M\otimes\sum_\ell\lambda_\ell\bra\ell,\nonumber\\
 B&=\sum_m B_m(I_2\otimes\bra m)
   =M^{-T}\otimes\sum_m c_{\ell_0m}\bra m.
 \label{eq:reassemble}
\end{align}
Defining $\bra a:=\sum_\ell\lambda_\ell\bra\ell$ and $\bra b:=\sum_m c_{\ell_0m}\bra m$ gives Eq.~\eqref{eq:reciprocal}. These are nonzero linear functionals on $\mathcal H_{A'}$ and $\mathcal H_{B'}$, respectively, and neither is required to be normalized.
\end{proof}

For a linear operator $X$, the Hilbert--Schmidt norm is $\|X\|_{\HS}:=\sqrt{\tr(X^\dagger X)}$.

\begin{lemma}
\label{lem:filter}
Let $M$ be an invertible qubit matrix and $L:=M\otimes M^{-T}$. Then
\begin{equation}
 L\ket\phi=\ket\phi,\qquad \bra\phi L=\bra\phi,
 \label{eq:leftright}
\end{equation}
and
\begin{equation}
 \|L\ket{\phi_\alpha}\|^2
 =\frac12\|MP_\alpha M^{-1}\|_{\HS}^2\ge1
 \quad(\alpha=x,y,z).
 \label{eq:errorbound}
\end{equation}
For every Bell-diagonal state $\rho$, the normalized state $L\rho L^\dagger/\tr(L\rho L^\dagger)$ has fidelity at most $F(\rho)$. If $\rho$ is full rank, equality holds if and only if $M$ is proportional to a unitary.
\end{lemma}

\begin{proof}
Applying Eq.~\eqref{eq:vecidentity} to $I_2$ gives
\begin{equation}
 L\ket\phi
 =\frac1{\sqrt2}\operatorname{vec}(MM^{-1})
 =\ket\phi.
 \label{eq:rightfixed}
\end{equation}
The left-vector identity must be verified independently because $L$ need not be unitary. The identity
\begin{equation}
 \bra\phi(R\otimes S)
 =\bra\phi[I_2\otimes(R^TS)]
 \label{eq:leftvec}
\end{equation}
gives $\bra\phi L=\bra\phi$, since $M^TM^{-T}=I_2$. This proves both statements in Eq.~\eqref{eq:leftright}.

For $\ket{\phi_\alpha}=(P_\alpha\otimes I_2)\ket\phi$, vectorization gives
\begin{equation}
 L\ket{\phi_\alpha}
 =\frac1{\sqrt2}\operatorname{vec}(MP_\alpha M^{-1}).
 \label{eq:errorvec}
\end{equation}
Since $\|\operatorname{vec}(X)\|^2=\tr(X^\dagger X)$, the equality in Eq.~\eqref{eq:errorbound} follows. Write $Q_\alpha:=MP_\alpha M^{-1}$ and let $s_1,s_2>0$ be its singular values. Similarity preserves determinants, and $|\det P_\alpha|=1$, so
\begin{equation}
 s_1s_2=|\det Q_\alpha|=1.
 \label{eq:singularproduct}
\end{equation}
Consequently,
\begin{equation}
 \frac12\|Q_\alpha\|_{\HS}^2
 =\frac{s_1^2+s_2^2}{2}
 \ge s_1s_2=1.
 \label{eq:singularbound}
\end{equation}
Equality holds precisely when both singular values equal one, or equivalently when $Q_\alpha$ is unitary.

For a Bell-diagonal state $\rho$, abbreviate its fixed-target fidelity as $F:=F(\rho)$ and write $\rho=F\Phi+\sum_\alpha q_\alpha\Phi_\alpha$, with $F+\sum_\alpha q_\alpha=1$. Equation~\eqref{eq:leftright} implies $L^\dagger\Phi L=\Phi$, so $\tr(\Phi L\rho L^\dagger)=F$. The trace of the unnormalized filtered state is
\begin{equation}
 D_\rho:=\tr(L\rho L^\dagger)
 =F+\frac12\sum_\alpha q_\alpha
          \|MP_\alpha M^{-1}\|_{\HS}^2.
 \label{eq:filtertrace}
\end{equation}
Using $F+\sum_\alpha q_\alpha=1$, we obtain
\begin{equation}
 D_\rho-1
 =\sum_\alpha q_\alpha
   \left(\frac12\|MP_\alpha M^{-1}\|_{\HS}^2-1\right)
 \ge0.
 \label{eq:filtertracegap}
\end{equation}
Since $L$ is invertible and $\rho$ is a state, $D_\rho>0$. The fidelity after normalization is therefore $F/D_\rho\le F$.

If $\rho$ is full rank, then $F>0$ and all $q_\alpha>0$. In that case $F/D_\rho=F$ requires $D_\rho=1$. Every summand in Eq.~\eqref{eq:filtertracegap} is nonnegative, so equality requires all three $Q_\alpha$ to be unitary. Put $H:=M^\dagger M>0$. Multiplying $Q_\alpha^\dagger Q_\alpha=I_2$ on the left by $M^\dagger$ and on the right by $M$ gives
\begin{equation}
 P_\alpha H P_\alpha=H,\qquad [H,P_\alpha]=0.
 \label{eq:Hcommute}
\end{equation}
Thus $H$ commutes with the single-qubit Pauli group in its irreducible representation on $\mathbb C^2$. By Schur's lemma~\cite{Etingof2011}, $H=hI_2$ with $h>0$ because $H$ is positive definite. The matrix $U:=M/\sqrt h$ is unitary, hence $M=\sqrt h\,U$. Conversely, if $M$ is proportional to a unitary, then $L=U\otimes U^*$ is unitary and preserves the target, so the fidelity is unchanged. The condition $F>0$ is needed for this equality argument: if $F=0$, the conditional fidelity remains zero for any invertible $M$.
\end{proof}

The operator $L$ in Lemma~\ref{lem:filter} is not assumed to be a physical Kraus operator by itself. Its trace in Eq.~\eqref{eq:filtertrace} may exceed one. The local contractions $M/\|M\|_\infty$ and $M^{-T}/\|M^{-T}\|_\infty$ implement a positive scalar multiple of $L$ as a filtering outcome. This scalar cancels in the normalized output. In the application below, the actual Kraus operator also includes the auxiliary functionals, and its success probability includes their contraction with $\tau$.

\begin{proof}[Proof of Theorem~\ref{thm:main}]
Choose an arbitrary finite product Kraus representation
$\E(X)=\sum_jE_jXE_j^\dagger$, with $E_j=A_j\otimes B_j$, and omit zero operators. Such a representation exists by the finite-dimensional Choi decomposition described in Sec.~\ref{sec:setting}.

First, we show that each product Kraus operator maps the subspace with first pair $\ket\phi$ into the target subspace of the output.
For $0<t<1$, write
\begin{equation}
 X_t:=\rho_t\otimes\tau,\qquad
 \rho_t=t\Phi+(1-t)\eta.
 \label{eq:testinputs}
\end{equation}
All four Bell coefficients of $\rho_t$ are positive, and $\tau>0$, so $X_t>0$. The nonzero positive operator $R:=\sum_jE_j^\dagger E_j$ therefore satisfies
\begin{equation}
 \tr\E(X_t)=\tr(RX_t)>0.
 \label{eq:positiveprob}
\end{equation}
For an individual Kraus term, the trace of its output is its probability weight, and the overlap of that output with $\Phi$ is its contribution to the unnormalized overlap with $\Phi$. Denote this output and these two weights by
\begin{align}
 O_j(t)&:=E_jX_tE_j^\dagger,\nonumber\\
 H_j(t)&:=\tr[\Phi O_j(t)],\qquad
 P_j(t):=\tr O_j(t).
 \label{eq:branchdefinitions}
\end{align}
When $P_j(t)>0$, the ratio $H_j(t)/P_j(t)$ is the conditional fidelity of this term. Summing the weights gives
\begin{align}
 p_{\E}(X_t)&=\sum_jP_j(t),\nonumber\\
 h_{\E}(X_t)&=\sum_jH_j(t).
 \label{eq:totalsdefinitions}
\end{align}
Equation~\eqref{eq:boundarytest} is thus $h_{\E}(X_{t_m})-t_mp_{\E}(X_{t_m})\ge0$.

The unnormalized quantities are continuous in $t$. Taking $t_m\to1$ gives
\begin{equation}
 \tr[\Phi\E(\Phi\otimes\tau)]
 \ge\tr\E(\Phi\otimes\tau).
 \label{eq:endpointineq}
\end{equation}
The reverse inequality holds because $0\le\Phi\le I$, so equality holds. Equivalently,
\begin{equation}
 \supp\E(\Phi\otimes\tau)\subseteq\mathbb C\ket\phi.
 \label{eq:puresupport}
\end{equation}
The support relation remains valid when $\E(\Phi\otimes\tau)=0$, so it does not require a conditional state to be defined at that input.

To apply the support restriction to individual Kraus operators, let $P_\perp:=I-\Phi$ be the output projection orthogonal to $\ket\phi$ and define the input isometry $J_\phi:\mathcal H_{A'B'}\to\mathcal H_{A_1B_1}\otimes\mathcal H_{A'B'}$ by $J_\phi\ket\chi:=\ket\phi\otimes\ket\chi$. Positivity gives
\begin{align}
 0&=\tr[P_\perp\E(\Phi\otimes\tau)]\nonumber\\
  &=\sum_j
    \|P_\perp E_jJ_\phi\tau^{1/2}\|_{\HS}^2.
 \label{eq:krauspositive}
\end{align}
Each squared norm vanishes. Since $\tau^{1/2}$ is invertible, $P_\perp E_jJ_\phi=0$ for every $j$. Every product Kraus operator therefore obeys Eq.~\eqref{eq:perfectconstraint}. By Lemma~\ref{lem:classify}, the indices split into $\mathcal A$, for which
$E_j=(M_j\otimes\bra{a_j})\otimes(M_j^{-T}\otimes\bra{b_j})$ with $M_j$ invertible, and $\mathcal Z$, for which $E_jJ_\phi=0$. Operators in $\mathcal A$ can succeed when the first pair is $\Phi$, whereas those in $\mathcal Z$ cannot.

Second, we calculate the fidelity contributions of the two types of operators. For $j\in\mathcal A$, put $L_j:=M_j\otimes M_j^{-T}$ and define $\ket{a_j}:=(\bra{a_j})^\dagger$ and $\ket{b_j}:=(\bra{b_j})^\dagger$. Write $\ket{a_jb_j}:=\ket{a_j}\otimes\ket{b_j}$, with neither vector required to be normalized. Because the first pair and the auxiliaries are independent, contraction of the auxiliary state gives
\begin{equation}
 E_j(\rho_t\otimes\tau)E_j^\dagger
 =\bra{a_jb_j}\tau\ket{a_jb_j}\,L_j\rho_tL_j^\dagger.
 \label{eq:activeoutput}
\end{equation}
Define $p_j:=\bra{a_jb_j}\tau\ket{a_jb_j}$. This number is positive because $\tau>0$ and the auxiliary functionals are nonzero. The target component in Eq.~\eqref{eq:activeoutput} contributes $tp_j\Phi$. For each $\Phi_\alpha$ component, the target overlap vanishes because
$\bra\phi L_j\ket{\phi_\alpha}=\langle\phi|\phi_\alpha\rangle=0$.
Its trace is multiplied by the squared norm in Eq.~\eqref{eq:errorbound}. The weighted sum of the trace factors for $\Phi_x$, $\Phi_y$, and $\Phi_z$ is
\begin{equation}
 c_j:=\frac12\sum_\alpha r_\alpha
          \|M_jP_\alpha M_j^{-1}\|_{\HS}^2\ge1,
 \label{eq:wc}
\end{equation}
where the inequality follows from Lemma~\ref{lem:filter}. Both $p_j$ and $c_j$ are independent of $t$. Thus
\begin{equation}
 H_j(t)=tp_j,\qquad
 P_j(t)=p_j[t+(1-t)c_j].
 \label{eq:activeprob}
\end{equation}
In particular, $p_j=P_j(1)$ is the weight on $\Phi\otimes\tau$, whereas it need not equal the weight at $t<1$. For each $j\in\mathcal A$, the conditional fidelity is
\begin{equation}
 \frac{H_j(t)}{P_j(t)}
 =\frac{t}{t+(1-t)c_j}\le t.
 \label{eq:branchfidelity}
\end{equation}
It cannot by itself improve the first-pair fidelity on these inputs.

For $j\in\mathcal Z$, define
\begin{equation}
 Y_j:=E_j(\eta\otimes\tau)E_j^\dagger,\qquad
 h_j:=\tr(\Phi Y_j),\quad \nu_j:=\tr Y_j.
 \label{eq:zeroY}
\end{equation}
These quantities refer to the fixed normalized input $\eta\otimes\tau$. Since $E_jJ_\phi=0$, the output on $X_t$ is $(1-t)Y_j$, and hence
\begin{equation}
 H_j(t)=(1-t)h_j,\qquad P_j(t)=(1-t)\nu_j.
 \label{eq:nullprob}
\end{equation}
We shall show that, for every $j\in\mathcal Z$,
\begin{equation}
 h_j<\nu_j.
 \label{eq:strict}
\end{equation}
This states that a nonzero operator of this type cannot produce only the target state from $\eta\otimes\tau$.

The operator $\eta+\Phi$ is diagonal in the Bell basis with eigenvalues $1,r_x,r_y,r_z>0$. Thus
\begin{equation}
 Q_0:=(\eta+\Phi)\otimes\tau>0.
 \label{eq:positiveQ}
\end{equation}
Adding the term annihilated by $E_j$ leaves $Y_j$ unchanged:
\begin{equation}
 Y_j=E_jQ_0E_j^\dagger.
 \label{eq:YwithQ}
\end{equation}
For any output vector $\ket u$, the quadratic form
$\bra uE_jQ_0E_j^\dagger\ket u=\|Q_0^{1/2}E_j^\dagger\ket u\|^2$
vanishes if and only if $E_j^\dagger\ket u=0$. Here $\ker T:=\{v:Tv=0\}$ denotes the null space of an operator $T$. Therefore
\begin{equation}
 \ker(E_jQ_0E_j^\dagger)
 =\ker E_j^\dagger
 =\ker(E_jE_j^\dagger).
 \label{eq:kernelidentity}
\end{equation}
Taking orthogonal complements, and using $E_jE_j^\dagger=(A_jA_j^\dagger)\otimes(B_jB_j^\dagger)$, gives
\begin{align}
 \supp Y_j&=\supp(E_jE_j^\dagger)\nonumber\\
 &=\supp(A_jA_j^\dagger)\otimes
   \supp(B_jB_j^\dagger).
 \label{eq:range}
\end{align}
All supports in this equation belong to the output spaces. The tensor-product identity follows by taking products of eigenvectors with positive eigenvalues of the two positive local operators.

Both local supports are nonzero because $E_j\ne0$. Their tensor product contains a nonzero product vector, which cannot be proportional to the Schmidt-rank-two vector $\ket\phi$. Hence $\supp Y_j$ is not contained in $\mathbb C\ket\phi$. It follows that
\begin{align}
 \nu_j-h_j
 &=\tr[(I-\Phi)Y_j]\nonumber\\
 &=\|(I-\Phi)Y_j^{1/2}\|_{\HS}^2>0.
 \label{eq:strictdifference}
\end{align}
This proves Eq.~\eqref{eq:strict}, including $\nu_j>0$. The argument does not assert that $Y_j$ is separable. The auxiliary state $\tau$ can contain entanglement, which a product Kraus operator may transfer to the output.

Third, we apply the fidelity inequality along $t_m\to1$ to exclude $\mathcal Z$ and require the local matrices in $\mathcal A$ to be proportional to unitaries. Combining the two classes gives, for every $0<t<1$,
\begin{align}
 h_{\E}(X_t)&=t\sum_{j\in\mathcal A}p_j
       +(1-t)\sum_{j\in\mathcal Z}h_j,\nonumber\\
 p_{\E}(X_t)&=\sum_{j\in\mathcal A}p_j[t+(1-t)c_j]
       +(1-t)\sum_{j\in\mathcal Z}\nu_j.
 \label{eq:explicittotals}
\end{align}
The corresponding fidelity difference is
\begin{equation}
 \Delta_{\E}(X_t;t)
 =h_{\E}(X_t)-tp_{\E}(X_t)
 =\sum_j[H_j(t)-tP_j(t)].
 \label{eq:sumuniversality}
\end{equation}
For $j\in\mathcal A$, Eq.~\eqref{eq:activeprob} gives
\begin{align}
 H_j(t)-tP_j(t)
 &=tp_j-tp_j[t+(1-t)c_j]\nonumber\\
 &=(1-t)tp_j(1-c_j).
 \label{eq:activegap}
\end{align}
For $j\in\mathcal Z$, Eq.~\eqref{eq:nullprob} gives
\begin{equation}
 H_j(t)-tP_j(t)
 =(1-t)(h_j-t\nu_j).
 \label{eq:nullgap}
\end{equation}
These are algebraic identities for all $0<t<1$. The nonnegative sign required by Eq.~\eqref{eq:boundarytest} is available at $t=t_m$. Dividing by $1-t_m>0$ at those values yields
\begin{equation}
 t_m\sum_{j\in\mathcal A}p_j(1-c_j)
 +\sum_{j\in\mathcal Z}(h_j-t_m\nu_j)\ge0.
 \label{eq:gap}
\end{equation}
Dividing by $1-t_m$ retains the first-order contribution as $t_m\to1$, whereas the undivided difference vanishes at $t=1$. The finite number of Kraus operators allows the limit to be taken term by term, giving
\begin{equation}
 \sum_{j\in\mathcal A}p_j(1-c_j)
 +\sum_{j\in\mathcal Z}(h_j-\nu_j)\ge0.
 \label{eq:limit}
\end{equation}
Every term in the first sum is nonpositive because $p_j>0$ and $c_j\ge1$. Every term in the second sum is strictly negative by Eq.~\eqref{eq:strict}. Thus
\begin{equation}
 \mathcal Z=\varnothing,\qquad c_j=1
 \quad\text{for all }j\in\mathcal A.
 \label{eq:equalityconclusion}
\end{equation}
For each remaining index $j\in\mathcal A$, Eqs.~\eqref{eq:wc} and~\eqref{eq:equalityconclusion} give
\begin{equation}
 0=c_j-1=\sum_\alpha r_\alpha
 \left(\frac12\|M_jP_\alpha M_j^{-1}\|_{\HS}^2-1\right).
 \label{eq:weightednormequality}
\end{equation}
Each term in parentheses is nonnegative by Lemma~\ref{lem:filter}, and every $r_\alpha$ is positive. Thus each term in parentheses is zero. The equality argument of Lemma~\ref{lem:filter} gives $M_j=s_jU_j$ with $s_j>0$ and $U_j$ unitary. Hence
\begin{equation}
 M_j^{-T}=s_j^{-1}U_j^{-T}=s_j^{-1}U_j^*.
 \label{eq:unitaryinverse}
\end{equation}
The scalars cancel between the two local factors of $E_j$, giving Eq.~\eqref{eq:rigidkraus}. This proves Theorem~\ref{thm:main}.
\end{proof}

The fidelity inequality in Eq.~\eqref{eq:boundarytest} is imposed only on the stated sequence with the same auxiliary state. No ordering of several input fidelities or condition on an entire family of states is assumed in Theorem~\ref{thm:main}. Section~\ref{sec:consequences} will show how ordered universality supplies these inputs.

We now calculate the output implied by Eq.~\eqref{eq:rigidkraus}. Let $\rho$ be any state of the first pair and let $\omega$ be any auxiliary state independent of $\rho$. Define
$p_j(\omega):=\bra{a_jb_j}\omega\ket{a_jb_j}$,
so that the fixed weight in the proof is $p_j=p_j(\tau)$, and put $V_j:=U_j\otimes U_j^*$. Direct contraction gives
\begin{equation}
 E_j(\rho\otimes\omega)E_j^\dagger
 =p_j(\omega)V_j\rho V_j^\dagger.
 \label{eq:contractedbranch}
\end{equation}
Summing these terms gives the output identity
\begin{equation}
 \E(\rho\otimes\omega)
 =\sum_jp_j(\omega)V_j\rho V_j^\dagger.
 \label{eq:rigidmap}
\end{equation}
The weights are probabilities of Kraus terms before conditioning on the aggregate success event. To check their normalization, put
$D_{\rm aux}:=\sum_j\ket{a_jb_j}\bra{a_jb_j}$.
After grouping the first pair and the auxiliary systems, trace nonincrease gives
\begin{equation}
 \sum_jE_j^\dagger E_j
 =I_{A_1B_1}\otimes D_{\rm aux}\le I,
 \qquad 0\le D_{\rm aux}\le I_{A'B'}.
 \label{eq:auxiliaryeffect}
\end{equation}
It follows that $0\le p_j(\omega)\le1$ and $\sum_jp_j(\omega)\le1$.

Because $V_j$ is unitary and fixes $\ket\phi$, Eq.~\eqref{eq:rigidmap} implies
\begin{align}
 \tr[\Phi\E(\rho\otimes\omega)]
 &=F(\rho)\sum_jp_j(\omega),\nonumber\\
 \tr\E(\rho\otimes\omega)
 &=\sum_jp_j(\omega).
 \label{eq:finalratio}
\end{align}
Whenever the second line is positive, their ratio is exactly $F(\rho)$. The normalized output is then a mixture of $V_j\rho V_j^\dagger$ with probabilities $p_j(\omega)/\sum_i p_i(\omega)$. These output formulas require independence between the first pair and the auxiliaries, but they require neither Bell diagonality of $\rho$ nor separability of $\omega$.

\section{No-go theorem for ordered universal purification}
\label{sec:consequences}

We apply Theorem~\ref{thm:main} to the product inputs of Definition~\ref{def:universal}. Under the segment condition of Definition~\ref{def:segment}, the first-pair state can approach $\Phi$ while the remaining pairs are held in the full-rank state $\sigma$. These allowed ordered inputs satisfy the hypothesis of Theorem~\ref{thm:main}. Its product Kraus representation then determines the output fidelity for every allowed joint input.

\begin{theorem}
\label{thm:ordered}
Let $n\ge2$ be finite and let $\Sset$ satisfy the segment condition in Definition~\ref{def:segment}. Then every nonzero separable $n$-to-1 success map universal with ordered fidelity on $\Sset$ preserves the highest input fidelity whenever it succeeds on an allowed input. The same holds for every convex complete Bell-diagonal family.
\end{theorem}

\begin{proof}
Write the full-rank state in the segment condition as
\begin{equation}
 \sigma=s\Phi+(1-s)\eta,\qquad 0<s<1.
 \label{eq:segmentdecomposition}
\end{equation}
Here $s:=F(\sigma)$ and $\eta=\sum_\alpha r_\alpha\Phi_\alpha$ has $r_\alpha>0$ and $\sum_\alpha r_\alpha=1$, because all Bell coefficients of $\sigma$ are positive. Identify $A':=A_2\cdots A_n$ and $B':=B_2\cdots B_n$, and choose the fixed auxiliary state $\tau:=\sigma^{\otimes(n-1)}>0$. For $s\le t\le1$,
\begin{align}
 \rho_t&=t\Phi+(1-t)\eta
        =(1-\lambda)\Phi+\lambda\sigma,\nonumber\\
 \lambda&:=\frac{1-t}{1-s},\qquad
 X_t:=\rho_t\otimes\sigma^{\otimes(n-1)}.
 \label{eq:orderedtests}
\end{align}
Since $0\le\lambda\le1$, the segment condition places every factor of $X_t$ in $\Sset$. Their fidelities are $t,s,\ldots,s$ in nonincreasing order. Universality on these inputs therefore requires
\begin{equation}
 \Delta_{\E}(X_t;t)
 =h_{\E}(X_t)-t p_{\E}(X_t)\ge0.
 \label{eq:universalitytotest}
\end{equation}
For example, the sequence
\begin{equation}
 t_m:=1-\frac{1-s}{m+1},\qquad m=1,2,\ldots,
 \label{eq:orderedsequence}
\end{equation}
satisfies $s<t_m<1$ and $t_m\to1$. Equation~\eqref{eq:universalitytotest} at these values is exactly the hypothesis of Theorem~\ref{thm:main}. The product Kraus representation supplied by that theorem, together with Eqs.~\eqref{eq:rigidmap} and~\eqref{eq:finalratio}, gives fidelity $F(\rho_1)$ for every allowed independent input whenever success has nonzero probability.

It remains to verify that convex complete Bell-diagonal families satisfy the segment condition. By completeness, for each $\alpha\in\{x,y,z\}$ choose a state $\rho^{(\alpha)}\in\Sset$ with a positive $\Phi_\alpha$ coefficient $q_\alpha^{(\alpha)}$. The three chosen states need not be distinct. Define
\begin{equation}
 \sigma:=\frac14\left[
 \Phi+\rho^{(x)}+\rho^{(y)}+\rho^{(z)}
 \right].
 \label{eq:convexsigma}
\end{equation}
Closure under binary convex combinations implies closure under this finite convex combination, so $\sigma\in\Sset$. Its target coefficient is positive because of the explicit $\Phi$ term, and its $\Phi_\alpha$ coefficient is at least $q_\alpha^{(\alpha)}/4>0$. Thus $\sigma$ is full rank. Convexity also gives
\begin{equation}
 \rho_\lambda:=(1-\lambda)\Phi+\lambda\sigma\in\Sset
 \quad(0\le\lambda\le1).
 \label{eq:convexsegment}
\end{equation}
These are the two requirements of the segment condition, so the already proved case applies. Since every LOCC success map is separable, the theorem applies in particular to LOCC.
\end{proof}

The exclusion of strict improvement does not require $\Phi$ itself to be an allowed input. More generally, it suffices that $\Sset$ contain a sequence
\begin{equation}
 \rho_{t_m}=t_m\Phi+(1-t_m)\eta\in\Sset,
 \qquad 0<t_m<1,\quad t_m\to1,
 \label{eq:noendpointsequence}
\end{equation}
where $\eta=\sum_\alpha r_\alpha\Phi_\alpha$ is fixed, with every $r_\alpha>0$ and $\sum_\alpha r_\alpha=1$. To see this, choose one member $\sigma:=\rho_{t_1}>0$ and hold the remaining pairs in $\tau:=\sigma^{\otimes(n-1)}$. All sufficiently large $m$ satisfy $t_m\ge t_1$, so $\rho_{t_m}\otimes\tau$ is an allowed ordered input. Theorem~\ref{thm:main} then gives the same fidelity equality on every allowed input. This argument uses only full-rank allowed states, and applies in particular after removing $\Phi$ from any family satisfying the segment condition.

An explicit example of the convex complete case is
\begin{equation}
 \Sset_{\rm iso}:=\{\omega_f:\tfrac34\le f\le1\}.
 \label{eq:convexexamplefamily}
\end{equation}
Mixing isotropic states averages $f$, so this set is convex. Every member is entangled, and it contains $\omega_1=\Phi$. By Eq.~\eqref{eq:isotropic}, the eigenvalues of $\omega_{3/4}$ are $3/4,1/12,1/12,1/12$, which verifies completeness and provides a full-rank member. Its segment to the target is
\begin{equation}
 (1-\lambda)\Phi+\lambda\omega_{3/4}
 =\omega_{1-\lambda/4},\qquad 0\le\lambda\le1.
 \label{eq:convexexamplesegment}
\end{equation}
For any fixed finite $n\ge2$, a separable success map universal on all independent choices $1\ge f_1\ge\cdots\ge f_n\ge3/4$ must therefore output fidelity $f_1$ whenever it succeeds. For instance, on
$\omega_{0.95}\otimes\omega_{0.85}\otimes\omega_{0.75}$,
every such nonzero map has positive success probability, since the input is full rank, and its output fidelity is exactly $0.95$. The requirement covers independently varying parameters $f_i$, rather than only inputs with identical factors.

For a fixed lower fidelity threshold $0\le F_*<1$, define the family
\begin{equation}
 \Sset_{F_*}:=\{\rho:F(\rho)\ge F_*\},
 \label{eq:threshold}
\end{equation}
where $\rho$ ranges over all two-qubit states.

\begin{corollary}
\label{cor:threshold}
For every fixed finite $n\ge2$, each nonzero separable $n$-to-1 success map universal with ordered fidelity on $\Sset_{F_*}$ preserves the highest input fidelity whenever it succeeds.
\end{corollary}

The same result holds when only Bell-diagonal states in $\Sset_{F_*}$ are allowed, when the lower bound is replaced by $F(\rho)>F_*$, or when both changes are made.

\begin{proof}
Choose $\max\{F_*,1/2\}<f<1$ and the full-rank isotropic state $\omega_f$. As in Eq.~\eqref{eq:convexexamplesegment}, its segment to $\Phi$ consists of isotropic states with fidelity $1-\lambda(1-f)\ge f>F_*$ for $0\le\lambda\le1$. Every state on the segment is entangled and Bell diagonal. Thus all four families described above contain the segment, and Theorem~\ref{thm:ordered} applies to each. Equation~\eqref{eq:rigidmap} establishes the fidelity equality also for allowed first-pair states that are not Bell diagonal.
\end{proof}

The same threshold results hold if the additional restriction $F(\rho)<1$ excludes $\Phi$: each resulting family still contains the sequence in Eq.~\eqref{eq:noendpointsequence}, for instance isotropic states with fidelities increasing from $f$ toward one.

Corollary~\ref{cor:threshold} excludes a strict universal improvement when the specified information consists of the fidelity ordering and a fixed lower threshold. The threshold can be arbitrarily close to one, and all auxiliary pairs used in the proof also exceed it. The statement holds for each finite $n$ and imposes no lower bound on success probability. It concerns a single map valid for independently varying inputs. Requiring identical copies restricts the allowed inputs in a different way, so recurrence purification of identical inputs~\cite{Bennett1996,Deutsch1996} is compatible with this result.

Retaining the first pair remains universal when its fidelity is known to be maximal. Without an ordering requirement, universality means $\Delta_{\E}(\rho_1\otimes\cdots\otimes\rho_n;\max_i F(\rho_i))\ge0$ for all $\rho_i\in\Sset$. Under this requirement, the segment condition excludes even a nonzero universal success map.

\begin{corollary}
\label{cor:unordered}
For any finite $n\ge2$, if $\Sset$ satisfies the segment condition in Definition~\ref{def:segment}, no nonzero separable $n$-to-1 success map is universal on all product inputs from $\Sset$ without an ordering restriction.
\end{corollary}

\begin{proof}
Assume, for contradiction, that such a map exists. Restricting its universality to the ordered inputs allows us to apply Theorem~\ref{thm:ordered}. Its proof yields Eq.~\eqref{eq:rigidmap} as an identity of the map on every independent first-pair and auxiliary input, including those whose fidelities are not in nonincreasing order.

Choose $s<t<1$, with $\sigma$ and $\rho_t$ as in Eqs.~\eqref{eq:segmentdecomposition} and~\eqref{eq:orderedtests}, and consider
\begin{equation}
 X:=\sigma\otimes\rho_t\otimes\sigma^{\otimes(n-2)}.
 \label{eq:unorderedtest}
\end{equation}
The last tensor factor is absent when $n=2$. The first-pair fidelity is $s$, while the highest input fidelity is $t>s$. The input does not satisfy Eq.~\eqref{eq:ordered}, but all factors of $X$ belong to $\Sset$. All factors are also full rank, so $X>0$ and the nonzero map succeeds with positive probability, by the argument in Eq.~\eqref{eq:positiveprob}.

Equation~\eqref{eq:rigidmap} applies because $X$ is independent across the first pair and its auxiliaries. It gives output fidelity $s$. Thus
\begin{equation}
 \tr[\Phi\E(X)]-t\tr\E(X)
 =(s-t)\tr\E(X)<0,
 \label{eq:unorderedcontradiction}
\end{equation}
contradicting universality relative to the highest input fidelity $t$.
\end{proof}

Using only full-rank factors in Eq.~\eqref{eq:unorderedtest} ensures that the success probability is positive. A success map could instead vanish on an input containing the rank-one state $\Phi$. Thus the contradiction uses positive success probability derived from $X>0$ and $\E\ne0$, without imposing it as an additional assumption.

The same contradiction applies to any family containing the sequence in Eq.~\eqref{eq:noendpointsequence}, even if $\Phi$ is excluded. Fix $\sigma=\rho_{t_1}$ and choose a sequence member with $t_m>t_1$. The resulting input $\sigma\otimes\rho_{t_m}\otimes\sigma^{\otimes(n-2)}$ is full rank and gives the strictly negative difference in Eq.~\eqref{eq:unorderedcontradiction}.

References~\cite{Zang2025,ZangThesis2026} leave open whether increasing the minimum input fidelity and allowing general separable or LOCC protocols can permit universality for the full two-qubit threshold family. The proof of Corollary~\ref{cor:threshold} establishes the segment condition for the full threshold families and for their Bell-diagonal restrictions, with either a strict or a non-strict lower fidelity bound. Corollary~\ref{cor:unordered} therefore answers this question negatively for every fixed finite number of independent input pairs: without fidelity ordering, no nonzero universal separable success map exists for any fixed threshold $0\le F_*<1$. Corollary~\ref{cor:threshold} further shows that supplying the fidelity ordering cannot permit a strict universal improvement. In that case, retaining the highest-fidelity pair is possible, but every universal separable success map must preserve its fidelity exactly whenever it succeeds. These statements do not classify every complete Bell-diagonal family, because completeness alone does not require states of fidelity below one to approach $\Phi$. Section~\ref{sec:counterexample} gives complete families for which a nontrivial universal LOCC protocol does exist.

\section{Ordered universal purification with entanglement catalysts}
\label{sec:catalysis}

Entanglement catalysis allows an additional shared state to assist a transformation while being returned unchanged~\cite{JonathanPlenio1999,EisertWilkens2000}. We consider whether this assistance can enable nontrivial ordered universal purification on the families studied in Sec.~\ref{sec:consequences}. For complete Bell-diagonal families, Theorem~11 of Ref.~\cite{Zang2025}, also presented as Theorem~7 in Ref.~\cite{ZangThesis2026}, excludes universal biCEP protocols without fidelity ordering, even with auxiliary Bell pairs that are returned unchanged. With fidelity ordering, the same result excludes nontrivial universal biCEP protocols under this assistance. Here the catalyst can be any finite-dimensional pure state, including an entangled state, and the operations can be arbitrary LOCC or separable success maps. Exact return is required only when the protocol succeeds.

\subsection{Pure catalysts and exact successful return}
\label{sec:catalystsetting}

For the definitions and reduction below, let $A:B$ and $\widehat A:\widehat B$ be arbitrary finite-dimensional input and output systems. The catalyst is shared on $C_A:C_B$. In the purification application, $A=A_1\cdots A_n$ and $B=B_1\cdots B_n$, and the output systems are qubits.

\begin{definition}
\label{def:catalyticassistance}
Fix a finite-dimensional pure state $C:=\ket c\bra c$ on $C_A C_B$, with $\langle c|c\rangle=1$, independent of the input state $X$ on $AB$. Let $\Lambda$ be a fixed LOCC or separable success map from $AC_A:BC_B$ to $\widehat A C_A:\widehat B C_B$. Its induced success map from $AB$ to $\widehat A\widehat B$ is
\begin{equation}
 \E_C(X):=\tr_{C_A C_B}\Lambda(X\otimes C),
 \qquad \E_C\ne0.
 \label{eq:inducedcatalytic}
\end{equation}
For an allowed input $X$ with $p_{\E_C}(X)>0$, the catalyst is returned exactly upon success if its reduced state, obtained by tracing out $\widehat A\widehat B$, satisfies
\begin{equation}
 \frac{\tr_{\widehat A\widehat B}\Lambda(X\otimes C)}
      {p_{\E_C}(X)}=C.
 \label{eq:exactcatalyticreturn}
\end{equation}
Exact pure catalytic assistance on a specified set of allowed inputs requires this condition whenever success has positive probability on an input in that set. No condition is imposed on the catalyst after failure.
\end{definition}

Here $p_{\E_C}(X)=\tr\E_C(X)=\tr\Lambda(X\otimes C)$ is the success probability. The unnormalized form of Eq.~\eqref{eq:exactcatalyticreturn} is
\begin{equation}
 \tr_{\widehat A\widehat B}\Lambda(X\otimes C)
 =p_{\E_C}(X)C.
 \label{eq:unnormalizedcatalyticreturn}
\end{equation}
This identity also holds when $p_{\E_C}(X)=0$, because a positive operator with zero trace is zero. The state $C$ and operation $\Lambda$ are fixed throughout the allowed input family. The exact-return condition applies to the combined successful output, without a separate return assumption for each Kraus operator. For ordered universal purification, the induced map $\E_C$ must satisfy Definition~\ref{def:universal} on the input pairs.

We will use the Schmidt rank of a pure vector~\cite{Horodecki2009}. For any nonzero vector $\ket v$ on a bipartition $L:R$, normalized or not, $\operatorname{SR}_{L:R}(\ket v)$ is the number of nonzero coefficients in its Schmidt decomposition. We omit the subscript when the bipartition is specified in the surrounding text. Multiplying the vector by a nonzero scalar does not change this number. In particular, write the normalized catalyst as
\begin{equation}
 \ket c=\sum_{\ell=1}^{r}\sqrt{\mu_\ell}
 \ket\ell_{C_A}\ket\ell_{C_B},
 \qquad \mu_\ell>0,\quad\sum_{\ell=1}^{r}\mu_\ell=1.
 \label{eq:catalystSchmidt}
\end{equation}
The catalyst has Schmidt rank $r\ge1$ and is entangled exactly when $r\ge2$. Local operations can change Schmidt coefficients when an outcome is selected~\cite{Vidal1999}, but a local linear operator cannot increase Schmidt rank~\cite{Horodecki2009,Vidal1999}. Schmidt ranks also multiply under tensor products of pure vectors. Both properties apply to the unnormalized Kraus outputs below, because Schmidt rank is independent of normalization.

\subsection{The induced operation is separable}
\label{sec:catalyststructure}

Tracing out a shared entangled state does not by itself imply that the induced operation is separable. The following lemma establishes separability from pure-state return on a single full-rank input.

\begin{lemma}
\label{lem:catalyst}
Let $C$, $\Lambda$, and $\E_C\ne0$ be as in Definition~\ref{def:catalyticassistance}. Suppose that for one full-rank state $X_0>0$ on $AB$,
\begin{equation}
 \tr_{\widehat A\widehat B}\Lambda(X_0\otimes C)
 =p_0C,\qquad p_0:=\tr\E_C(X_0).
 \label{eq:oneinputreturn}
\end{equation}
Then $\E_C$ is a separable success map, and for every input operator $X$,
\begin{equation}
 \Lambda(X\otimes C)=\E_C(X)\otimes C.
 \label{eq:globalreturn}
\end{equation}
\end{lemma}

\begin{proof}
Every LOCC success map is separable. Thus, in either operation class in the statement, we can write
\begin{equation}
 \Lambda(Z)=\sum_j L_j ZL_j^\dagger,
 \qquad L_j=\mathcal A_j\otimes\mathcal B_j,
 \label{eq:catalyticKraus}
\end{equation}
with finitely many terms. Here $\mathcal A_j:AC_A\to\widehat A C_A$ and $\mathcal B_j:BC_B\to\widehat B C_B$. The induced map is completely positive as a composition of preparation of $C$, $\Lambda$, and a partial trace. All tensor products below are understood with the corresponding physical subsystems grouped in the indicated bipartitions.

\emph{First, pure-state return and full rank constrain the image of every Kraus operator.} Let $\epsilon:=\lambda_{\min}(X_0)>0$ denote the smallest eigenvalue of $X_0$. Since $X_0\ge\epsilon I_{AB}$, complete positivity and $\E_C\ne0$ give
\begin{equation}
 p_0\ge\epsilon\tr\E_C(I_{AB})>0.
 \label{eq:catalystpositive}
\end{equation}
The final inequality follows because a nonzero completely positive map has at least one nonzero Kraus operator, so its value on the identity is a nonzero positive operator.

Introduce the input and output isometries
\begin{equation}
 J_{\rm in}\ket v:=\ket v\otimes\ket c,
 \qquad J_{\rm out}\ket u:=\ket u\otimes\ket c,
 \label{eq:catalystisometries}
\end{equation}
where $\ket v$ belongs to $AB$ and $\ket u$ belongs to $\widehat A\widehat B$. They satisfy $J_{\rm in}^\dagger J_{\rm in}=I_{AB}$ and $J_{\rm out}^\dagger J_{\rm out}=I_{\widehat A\widehat B}$. The projection onto the output subspace orthogonal to the returned catalyst is
\begin{align}
 P_c^\perp&:=I_{\widehat A\widehat B}\otimes(I_{C_A C_B}-C)\nonumber\\
 &=I-J_{\rm out}J_{\rm out}^\dagger.
 \label{eq:catalystprojection}
\end{align}
Set $\Omega_0:=\Lambda(X_0\otimes C)\ge0$. Taking the trace of Eq.~\eqref{eq:oneinputreturn} against $I-C$ yields
\begin{equation}
 0=\tr(P_c^\perp\Omega_0)
 =\|P_c^\perp\Omega_0^{1/2}\|_\HS^2.
 \label{eq:pureoutputsupp}
\end{equation}
Hence $\Omega_0$ is supported on the range of $J_{\rm out}$. It therefore has the form $S\otimes C$ for a positive operator $S$ on $\widehat A\widehat B$. Taking the partial trace over $C_A C_B$ identifies $S=\E_C(X_0)$, proving
\begin{equation}
 \Lambda(X_0\otimes C)=\E_C(X_0)\otimes C.
 \label{eq:purereturnproduct}
\end{equation}
Thus a pure returned reduced state already excludes correlations between the output systems and catalyst on this input.

Since $X_0\otimes C=J_{\rm in}X_0J_{\rm in}^\dagger$, the same zero trace can be expanded using Eq.~\eqref{eq:catalyticKraus}:
\begin{align}
 0&=\tr[P_c^\perp\Lambda(X_0\otimes C)]\nonumber\\
  &=\sum_j\|P_c^\perp L_jJ_{\rm in}X_0^{1/2}\|_\HS^2.
 \label{eq:catalystsupportsum}
\end{align}
Each summand is nonnegative and must vanish. Invertibility of $X_0^{1/2}$ then gives
\begin{equation}
 P_c^\perp L_jJ_{\rm in}=0.
 \label{eq:catalystbranchsupport}
\end{equation}
This holds for every $j$ and extends return on one full-rank state to a constraint on every input vector.

\emph{Second, each Kraus operator defines a unique induced linear operator.} Equation~\eqref{eq:catalystbranchsupport} implies that the range of $L_jJ_{\rm in}$ lies in the range of $J_{\rm out}$. Define
\begin{equation}
 T_j:=J_{\rm out}^\dagger L_jJ_{\rm in}.
 \label{eq:inducedoperatorformula}
\end{equation}
Using $J_{\rm out}J_{\rm out}^\dagger=I-P_c^\perp$ gives
\begin{align}
 L_jJ_{\rm in}&=J_{\rm out}T_j,\nonumber\\
 L_j(\ket v\otimes\ket c)&=T_j\ket v\otimes\ket c.
 \label{eq:branchcancellation}
\end{align}
The second equality holds for every vector $\ket v$ on $AB$. The operator $T_j:AB\to\widehat A\widehat B$ is unique: multiplying $L_jJ_{\rm in}=J_{\rm out}T_j$ by $J_{\rm out}^\dagger$ recovers Eq.~\eqref{eq:inducedoperatorformula}.
For every operator $X$ on $AB$, substitution into the Kraus sum gives
\begin{align}
 \Lambda(X\otimes C)
 &=J_{\rm out}\left(\sum_jT_jXT_j^\dagger\right)
   J_{\rm out}^\dagger\nonumber\\
 &=\left(\sum_jT_jXT_j^\dagger\right)\otimes C.
 \label{eq:factorizedKrausoutput}
\end{align}
Taking the partial trace over the catalyst proves Eq.~\eqref{eq:globalreturn} and supplies a Kraus representation of the induced map:
\begin{equation}
 \E_C(X)=\sum_jT_jXT_j^\dagger.
 \label{eq:inducedKraus}
\end{equation}
This map is completely positive. It is trace nonincreasing because, for $X\ge0$,
\begin{equation}
 \tr\E_C(X)=\tr\Lambda(X\otimes C)
 \le\tr(X\otimes C)=\tr X.
 \label{eq:inducedtracenonincreasing}
\end{equation}
To establish separability, it remains to show that every operator $T_j$ factors across $A:B$.

\emph{Third, local reference systems force each nonzero induced operator to be a product operator.} Let $d_A:=\dim\mathcal H_A$ and $d_B:=\dim\mathcal H_B$, and use the reference systems $R_A,R_B$ of Eq.~\eqref{eq:choi}. Normalize the local reference vectors by defining
\begin{align}
 \ket{\chi_A}&:=\frac{\ket{\Omega_A}}{\sqrt{d_A}}
   =\frac1{\sqrt{d_A}}
    \sum_{a=1}^{d_A}\ket a_A\ket a_{R_A},\nonumber\\
 \ket{\chi_B}&:=\frac{\ket{\Omega_B}}{\sqrt{d_B}}
   =\frac1{\sqrt{d_B}}
    \sum_{b=1}^{d_B}\ket b_B\ket b_{R_B}.
 \label{eq:localreferences}
\end{align}
The tensor product of these two locally prepared states is a product vector across Alice's and Bob's laboratories. Across $AR_AC_A:BR_BC_B$, the combined vector therefore satisfies
\begin{equation}
 \operatorname{SR}(\ket{\chi_A}\ket{\chi_B}\ket c)=r.
 \label{eq:referenceinputrank}
\end{equation}
For $T_j\ne0$, define the unnormalized vector
\begin{equation}
 \ket{\Gamma_j}:=(T_j\otimes I_{R_AR_B})
    \ket{\chi_A}\ket{\chi_B}.
 \label{eq:ChoiVector}
\end{equation}
This is the vector associated with $T_j$ by the operator--state correspondence in Eq.~\eqref{eq:choiproduct}, scaled by $1/\sqrt{d_Ad_B}$ because the reference vectors here are normalized. It is nonzero because $\|\Gamma_j\|^2=\tr(T_j^\dagger T_j)/(d_Ad_B)>0$. Applying Eq.~\eqref{eq:branchcancellation} to each component of the reference states shows that the corresponding output vector is $\ket{\Gamma_j}\otimes\ket c$.

The acting operator has local factors $\mathcal A_j\otimes I_{R_A}$ and $\mathcal B_j\otimes I_{R_B}$, so it cannot increase the Schmidt rank~\cite{Horodecki2009,Vidal1999}. The Schmidt rank of $\ket{\Gamma_j}\otimes\ket c$ is the product of the Schmidt ranks of its two factors. Taking the Schmidt rank of $\ket{\Gamma_j}$ across $\widehat A R_A:\widehat B R_B$ and that of the full output across $\widehat A R_AC_A:\widehat B R_BC_B$, we obtain
\begin{equation}
 r\,\operatorname{SR}(\ket{\Gamma_j})
 =\operatorname{SR}(\ket{\Gamma_j}\otimes\ket c)\le r.
 \label{eq:rankcancellation}
\end{equation}
Since $0<r<\infty$ and $\ket{\Gamma_j}\ne0$, its Schmidt rank is one. Write $\ket{\Gamma_j}=\ket{u_j}_{\widehat A R_A}\otimes\ket{v_j}_{\widehat B R_B}$. In fixed product bases of $\widehat A R_A$ and $\widehat B R_B$, respectively,
\begin{equation}
 \ket{\Gamma_j}
 =\frac1{\sqrt{d_Ad_B}}
   \sum_{u,v,a,b}(T_j)_{uv,ab}
     \ket{ua}\ket{vb}.
 \label{eq:ChoiCoefficients}
\end{equation}
Here $u$ and $v$ label orthonormal bases of $\mathcal H_{\widehat A}$ and $\mathcal H_{\widehat B}$, while $a$ and $b$ label the input bases and their corresponding reference bases. Comparing coefficients gives $(T_j)_{uv,ab}=\sqrt{d_Ad_B}(u_j)_{ua}(v_j)_{vb}$. Define $(\widetilde A_j)_{ua}:=\sqrt{d_A}(u_j)_{ua}$ and $(\widetilde B_j)_{vb}:=\sqrt{d_B}(v_j)_{vb}$. Then
\begin{equation}
 T_j=\widetilde A_j\otimes\widetilde B_j.
 \label{eq:inducedproduct}
\end{equation}
Omitting zero $T_j$ from Eq.~\eqref{eq:inducedKraus} gives a product Kraus representation of $\E_C$. Together with Eq.~\eqref{eq:inducedtracenonincreasing}, this proves that $\E_C$ is a separable success map.
\end{proof}

Full rank and purity have distinct roles in this proof. Full rank of $X_0$ gives Eq.~\eqref{eq:catalystbranchsupport} on the entire input space. Purity fixes the returned catalyst factor for each successful Kraus contribution, enabling cancellation of its finite Schmidt rank. The lemma therefore does not exclude catalytic transformations on rank-deficient input states~\cite{JonathanPlenio1999,EisertWilkens2000}.

\subsection{Application to ordered universal purification}

We now apply the separability result of Lemma~\ref{lem:catalyst} to ordered purification. The segment condition supplies the required full-rank input, after which Theorem~\ref{thm:ordered}, derived from Theorem~\ref{thm:main}, excludes strict universal improvement.

\begin{theorem}
\label{thm:catalyticpurification}
Let $n\ge2$ be finite and let $\Sset$ satisfy the segment condition in Definition~\ref{def:segment}. Let $\E_C\ne0$ be induced by exact pure catalytic assistance as in Definition~\ref{def:catalyticassistance}. If $\E_C$ is universal with ordered fidelity on $\Sset$, it preserves the highest input fidelity whenever it succeeds on an allowed input.
\end{theorem}

\begin{proof}
Let $\sigma>0$ be the Bell-diagonal state in the segment condition and set $X_0:=\sigma^{\otimes n}$. All its factors belong to $\Sset$ and have equal fidelity, so $X_0$ is an allowed ordered input. It is full rank. Since $\E_C\ne0$, Eq.~\eqref{eq:catalystpositive} gives $p_{\E_C}(X_0)>0$, and exact successful return supplies Eq.~\eqref{eq:oneinputreturn}. Lemma~\ref{lem:catalyst} therefore makes $\E_C$ a separable success map.

By hypothesis, $\E_C$ is universal with ordered fidelity on $\Sset$. Theorem~\ref{thm:ordered} therefore applies to this separable success map. Its proof uses Theorem~\ref{thm:main} to obtain Eq.~\eqref{eq:rigidmap}, whose correlated local unitaries preserve the overlap with $\Phi$. Consequently, for every allowed input $X=\rho_1\otimes\cdots\otimes\rho_n$,
\begin{equation}
 h_{\E_C}(X)=F(\rho_1)p_{\E_C}(X).
 \label{eq:catalyticfidelityequality}
\end{equation}
This is fidelity preservation on the whole allowed domain, including inputs with zero success probability. Whenever $p_{\E_C}(X)>0$, normalization gives $F_{\E_C}(X)=F(\rho_1)$.
\end{proof}

Exact pure catalytic assistance therefore cannot provide a strict universal improvement on these families. The result includes the convex complete Bell-diagonal families and the arbitrarily high fidelity threshold families of Sec.~\ref{sec:consequences}. It uses no positive lower bound on success probability, places no restriction on the finite Schmidt spectrum of the catalyst, and requires no return after failure. The same argument covers families containing the sequence in Eq.~\eqref{eq:noendpointsequence}, without requiring $\Phi$ to be allowed: any sequence member supplies $\sigma>0$ and hence $X_0=\sigma^{\otimes n}>0$. Applying Lemma~\ref{lem:catalyst} before Corollary~\ref{cor:unordered} also extends the unordered impossibility result to exact pure catalytic assistance.
\section{Complete families admitting universal LOCC purification}
\label{sec:counterexample}

In Sec.~\ref{sec:consequences}, convexity and completeness jointly imply the segment condition. Completeness alone does not provide this condition. Universal purification is already possible on some incomplete families. For example, the fixed recurrence protocol never lowers the higher input fidelity when both pairs have states of the form $f\Phi+(1-f)\Phi_x$ with independently chosen $f\in(1/2,1]$~\cite{ZangNetworks2025}. This family has only one Bell component orthogonal to $\Phi$ and is therefore incomplete.

We now construct complete finite and infinite families, each containing $\Phi$ and full-rank states, on which a fixed 3-to-1 LOCC protocol performs nontrivial universal purification with ordered fidelity. The protocol combines retention of the first pair with a standard recurrence operation on the other two pairs. Its success rule ensures that the recurrence contribution vanishes when the first pair is $\Phi$. Together with the restrictions on the allowed fidelities, this rule gives the required guarantee for every allowed ordered input. The constructions establish that completeness alone cannot extend the specified biCEP no-go theorem~\cite{Zang2025} to general LOCC.

\subsection{A finite complete family}

Fix $1/2<f<1$ and consider
\begin{equation}
 \mathcal T_f:=\{\Phi,\omega_f\},
 \label{eq:twostate}
\end{equation}
where $\omega_f$ is the isotropic state defined in Eq.~\eqref{eq:isotropic}. Both members are entangled, the family contains $\Phi$, and $\omega_f$ has positive coefficients of all three $\Phi_\alpha$. Thus $\mathcal T_f$ is complete. Its only fidelities are $f$ and one, so it contains no segment joining $\Phi$ to a full-rank state.

\begin{proposition}
\label{prop:counter}
For every fixed $1/2<f<1$, there exists a 3-to-1 LOCC protocol that is universal with ordered fidelity on $\mathcal T_f$, succeeds with probability at least $1/2$ on every allowed input, and strictly increases the highest input fidelity on $\omega_f^{\otimes3}$.
\end{proposition}
\begin{proof}
Let the input be $X:=\rho_1\otimes\rho_2\otimes\rho_3$, with each $\rho_i\in\mathcal T_f$ and $F(\rho_1)\ge F(\rho_2)\ge F(\rho_3)$. Alice holds $A_1,A_2,A_3$ and Bob holds $B_1,B_2,B_3$. They use the following protocol.
\begin{enumerate}
 \item Alice chooses a uniformly random bit $\xi\in\{0,1\}$ and sends it to Bob. If $\xi=0$, they retain pair~1, discard pairs~2 and~3, and declare success. If $\xi=1$, they continue to step~2.
 \item Alice measures $A_1$ and Bob measures $B_1$, each in the computational basis $\{\ket0,\ket1\}$. They record the classical outcomes $u_1,v_1\in\{0,1\}$, communicate them, and discard the measured pair. If $u_1=v_1$, they declare failure and stop. If $u_1\ne v_1$, they continue to step~3.
 \item Alice applies a controlled-NOT (CNOT) gate from $A_2$ to $A_3$, and Bob applies one from $B_2$ to $B_3$. They then measure $A_3$ and $B_3$ separately in their computational bases, communicate the outcomes $u_3,v_3$, and discard pair~3. If $u_3=v_3$, they retain pair~2 and declare success. If $u_3\ne v_3$, they declare failure.
 \item On success, local swaps transfer the retained pair to the designated output systems $\widehat A\widehat B$. On failure, the parties discard all remaining input systems and report the failure outcome.
\end{enumerate}
Every quantum operation is local, and all communication is classical. The protocol therefore defines an LOCC instrument. Let $\E$ denote its success map after the retained pair has been transferred to $\widehat A\widehat B$ and all successful classical records have been discarded. The fidelity requirement concerns the output conditioned on this combined success event. The protocol is independent of $f$, and the choice $\xi=0$ alone gives success probability $1/2$ on every input.

To verify ordered universality, we evaluate $\Delta_{\E}(X;F(\rho_1))$ from Eq.~\eqref{eq:fidelitydifference}. Its sign is the sign of the conditional fidelity gain because the success probability is positive. The branch $\xi=0$ retains pair~1 and therefore contributes zero to this difference. The contribution from $\xi=1$ is determined by the following two cases.

\emph{Case 1: $\rho_1=\Phi$.} Since $\ket\phi=(\ket{00}+\ket{11})/\sqrt2$, the outcomes in step~2 are $00$ or $11$, each with probability $1/2$ conditional on $\xi=1$. Both give failure. Thus only $\xi=0$ contributes to the successful output, and
\begin{equation}
 \E(\Phi\otimes\rho_2\otimes\rho_3)=\tfrac12\Phi,
 \qquad \Delta_{\E}(X;1)=0.
 \label{eq:perfectcounter}
\end{equation}
The success probability is $1/2$, and the normalized output is $\Phi$. This conclusion holds for every allowed choice of $\rho_2$ and $\rho_3$.

\emph{Case 2: $\rho_1=\omega_f$.} The ordering and the membership condition $\rho_i\in\mathcal T_f$ force $\rho_2=\rho_3=\omega_f$: a later pair in state $\Phi$ would have larger fidelity than the first. We first calculate the probability of the unequal outcomes in step~2, and then the unnormalized output of step~3 for its two successful measurement outcomes.

In the ordered basis $\{\ket{00},\ket{01},\ket{10},\ket{11}\}$ of a pair, the state of each pair is
\begin{equation}
 \omega_f=
 \begin{pmatrix}
 a&0&0&c\\
 0&e&0&0\\
 0&0&e&0\\
 c&0&0&a
 \end{pmatrix},
 \label{eq:omegamatrix}
\end{equation}
where the entries are
\begin{equation}
 e:=\frac{1-f}{3},\qquad a:=\frac{f+e}{2},\qquad c:=\frac{f-e}{2}.
 \label{eq:ace}
\end{equation}
The outcomes $(u_1,v_1)=(0,1)$ and $(1,0)$ each have probability $e$. Their total probability, conditional on $\xi=1$, is therefore
\begin{equation}
 q(f):=2e=\frac{2(1-f)}3.
 \label{eq:detector}
\end{equation}
Conditioning on either of these outcomes leaves pairs~2 and~3 in $\omega_f\otimes\omega_f$, because the three input pairs are independent. Accordingly, the probability of reaching step~3, including the initial random choice, is $q(f)/2$.

Step~3 is the bilateral-CNOT recurrence operation~\cite{Bennett1996,Duer2007}. Write its unitary as
\begin{equation}
 C_{2\to3}:=\operatorname{CNOT}_{A_2\to A_3}
 \otimes\operatorname{CNOT}_{B_2\to B_3}.
 \label{eq:bilateralcnot}
\end{equation}
Using the pair-ordered computational basis $\ket{ik}_{A_2B_2}\ket{jl}_{A_3B_3}$, this operator acts as
\begin{equation}
 \ket{ik}\ket{jl}
 \longmapsto
 \ket{ik}\ket{j\mathbin\oplus i,\,l\mathbin\oplus k},
 \label{eq:cnotbasis}
\end{equation}
where all indices are bits and $\oplus$ denotes addition modulo two. The two successful measurement outcomes on pair~3 are $00$ and $11$. For outcome $mm$, the operator mapping the two input pairs to the retained pair is
\begin{equation}
 W_m:=(I_{A_2B_2}\otimes\bra{mm}_{A_3B_3})C_{2\to3},
 \qquad m=0,1.
 \label{eq:recurrencekraus}
\end{equation}
After discarding which of these two outcomes occurred, the recurrence success operation on a two-pair operator $Z$ is
\begin{equation}
 \mathcal R(Z):=\sum_{m=0}^1 W_mZW_m^\dagger.
 \label{eq:recurrencemap}
\end{equation}
For the present input, define its unnormalized successful output by
\begin{equation}
 R_f:=\mathcal R(\omega_f\otimes\omega_f).
 \label{eq:Rdefinition}
\end{equation}
The outcomes $00$ and $11$ are recorded separately, so the corresponding unnormalized outputs are added incoherently in this sum.

The basis action in Eq.~\eqref{eq:cnotbasis} determines the matrix elements of $R_f$. Obtaining $mm$ requires $j=m\oplus i$ and $l=m\oplus k$, so
\begin{align}
 \bra{ik}R_f\ket{i'k'}
 =\sum_{m=0}^1
 &\bra{ik,\,m\oplus i,\,m\oplus k}
 (\omega_f\otimes\omega_f)\nonumber\\
 &\times\ket{i'k',\,m\oplus i',\,m\oplus k'}.
 \label{eq:Rentries}
\end{align}
For example, the $00,00$ entry is $a^2+a^2$, the $01,01$ entry is $e^2+e^2$, and the $00,11$ entry is $c^2+c^2$. The corresponding symmetric entries have the same values, and the remaining entries vanish. Hence the unnormalized output on the retained pair is
\begin{equation}
 R_f=
 \begin{pmatrix}
 2a^2&0&0&2c^2\\
 0&2e^2&0&0\\
 0&0&2e^2&0\\
 2c^2&0&0&2a^2
 \end{pmatrix}.
 \label{eq:Rmatrix}
\end{equation}

We next extract the success probability and fidelity of this recurrence operation, conditional on having reached step~3. Its success probability is the trace of $R_f$:
\begin{align}
 D(f)&:=\tr R_f=4(a^2+e^2)\nonumber\\
 &=f^2+\frac{2f(1-f)}3+\frac{5(1-f)^2}{9}.
 \label{eq:recurrenceD}
\end{align}
To obtain the fidelity of the normalized state $R_f/D(f)$, we also need its unnormalized overlap with $\Phi$. Define
\begin{align}
 N(f)&:=\tr(\Phi R_f)=2(a^2+c^2)\nonumber\\
 &=f^2+\frac{(1-f)^2}{9}.
 \label{eq:recurrenceN}
\end{align}
The fidelity conditioned on the success of step~3 is consequently
\begin{equation}
 G(f):=F\!\left(\frac{R_f}{D(f)}\right)=\frac{N(f)}{D(f)}.
 \label{eq:recurrence}
\end{equation}
The function $G(f)$ is the recurrence fidelity derived in Ref.~\cite{Bennett1996}. The quantities in Eqs.~\eqref{eq:recurrenceD}--\eqref{eq:recurrence} concern pairs~2 and~3 conditional on reaching step~3. Since $D(f)>0$, the comparison with the common fidelity $f$ of the input pairs is
$G(f)-f=[N(f)-fD(f)]/D(f)$. Substitution from Eqs.~\eqref{eq:recurrenceD} and~\eqref{eq:recurrenceN} gives
\begin{equation}
 N(f)-fD(f)=\frac{(1-f)(2f-1)(4f-1)}9>0.
 \label{eq:strictgain}
\end{equation}
Thus the recurrence output has fidelity greater than $f$ for every $1/2<f<1$.

It remains to include both successful branches of the three-pair protocol. The branch $\xi=0$ outputs $\omega_f$ with probability $1/2$. The branch $\xi=1$ reaches step~3 with probability $q(f)/2$ and contributes its unnormalized output $R_f$. After routing either retained pair to the same output systems, their contributions add to
\begin{equation}
 \E(\omega_f^{\otimes3})=
 \frac12\omega_f+\frac12q(f)R_f.
 \label{eq:fullmatrix}
\end{equation}
Equation~\eqref{eq:fidelitydifference} therefore gives
\begin{equation}
 \Delta_{\E}(\omega_f^{\otimes3};f)
 =\frac{q(f)}2\bigl[N(f)-fD(f)\bigr]>0.
 \label{eq:countergapnoisy}
\end{equation}
Here $q(f)>0$, and the bracket is positive by Eq.~\eqref{eq:strictgain}. This establishes the strict improvement required for nontrivial purification. The total probability that the protocol succeeds is the trace of Eq.~\eqref{eq:fullmatrix}:
\begin{align}
 p_{\rm succ}(f)&:=p_{\E}(\omega_f^{\otimes3})\nonumber\\
 &=\frac{1+q(f)D(f)}2.
 \label{eq:counterprobability}
\end{align}
The fidelity with $\Phi$ conditioned on the combined success event is
\begin{align}
 F_{\rm out}(f)&:=F_{\E}(\omega_f^{\otimes3})\nonumber\\
 &=\frac{f+q(f)N(f)}{1+q(f)D(f)}.
 \label{eq:counterfidelity}
\end{align}
Subtracting the first-pair fidelity $f$ expresses the improvement after both successful branches have been included:
\begin{equation}
 F_{\rm out}(f)-f
 =\frac{q(f)[N(f)-fD(f)]}{1+q(f)D(f)}>0.
 \label{eq:totalcounter}
\end{equation}

The allowed ordered product inputs are exactly
\begin{equation}
 \begin{gathered}
 \Phi^{\otimes3},\quad
 \Phi^{\otimes2}\otimes\omega_f,\\
 \Phi\otimes\omega_f^{\otimes2},\quad
 \omega_f^{\otimes3}.
 \end{gathered}
 \label{eq:allowedtwostate}
\end{equation}
Case~1 gives equality in the universality condition for the first three inputs, and Case~2 gives strict inequality for the last. Together with the success probability of at least $1/2$, this proves the proposition.
\end{proof}

\begin{figure}[t]
 \centering
 \includegraphics[width=\columnwidth]{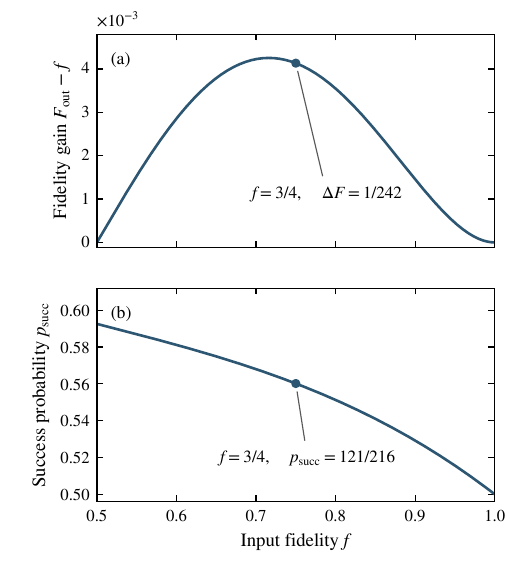}
 \caption{Performance of the 3-to-1 LOCC protocol on the input $\omega_f^{\otimes3}$. (a) Conditional fidelity gain $\Delta F:=F_{\rm out}(f)-f$ from Eq.~\eqref{eq:totalcounter}. (b) Total success probability $p_{\rm succ}(f)$ from Eq.~\eqref{eq:counterprobability}. The marked point at $f=3/4$ has gain $1/242$ and success probability $121/216$. For the allowed inputs whose first pair is in state $\Phi$, the conditional fidelity is one and the success probability is $1/2$. Each value of $f$ specifies a separate input family $\mathcal T_f$.}
 \label{fig:counter}
\end{figure}

The size of the fidelity gain and its associated success probability can be illustrated on the specific input $\omega_{3/4}^{\otimes3}$. Equations~\eqref{eq:counterprobability} and~\eqref{eq:counterfidelity} yield
\begin{equation}
 p_{\rm succ}=\frac{121}{216},\qquad
 F_{\rm out}=\frac{365}{484}=\frac34+\frac1{242}.
 \label{eq:exactexample}
\end{equation}
Figure~\ref{fig:counter} shows the fidelity gain and success probability as the common input fidelity varies. For the other three allowed ordered inputs from $\mathcal T_{3/4}$, Eq.~\eqref{eq:perfectcounter} instead gives success probability $1/2$ and conditional fidelity one.

The circuit is independent of $f$, but the universality statement is made separately for each fixed family $\mathcal T_f$. Taking their union enlarges the allowed input family by allowing different isotropic fidelities in the same product input. For instance, on $\omega_{9/10}\otimes\omega_{3/5}\otimes\omega_{3/5}$ the event $u_1\ne v_1$ has positive probability, whereas the recurrence output has fidelity $G(3/5)=85/137<9/10$. Mixing that successful contribution with the retained first pair strictly lowers the conditional fidelity. Thus the same circuit is not universal on the union of all these families.

The construction demonstrates a difference between the specified biCEP class and general LOCC under the ordered universality requirement. In the biCEP class of Ref.~\cite{Zang2025}, a fixed $n$-qubit Clifford $C$ and its complex conjugate $C^*$ act on Alice's and Bob's systems, respectively. The last $n-1$ pairs are then measured, the prescribed first output pair is retained, and success requires matching local outcomes on every measured pair. The present protocol uses local Clifford gates and computational-basis measurements, but its rule for success and its choice of retained input pair depend on $\xi$: it either retains pair~1 unconditionally or requires unequal outcomes on pair~1 and equal outcomes on pair~3 before retaining pair~2. These rules are not those of the specified biCEP class. Moreover, the present protocol's strict fidelity improvement on a complete family excludes an implementation within that class by the biCEP no-go theorem~\cite{Zang2025}. Proposition~\ref{prop:counter} therefore establishes that completeness alone cannot imply the corresponding impossibility under general LOCC.

For any $n\ge3$, applying the protocol to the first three pairs and discarding the others gives an $n$-to-1 counterexample on the same family. Ordering ensures that the first of those three pairs still has the highest input fidelity.

\subsection{An infinite complete family}

The same 3-to-1 protocol also works when the allowed fidelities vary continuously over a suitable interval. For fidelity bounds $1/2<f_-<f_+<1$, define
\begin{equation}
 \mathcal T_{f_-,f_+}:=\{\Phi\}\cup\{\omega_f:f_-\le f\le f_+\}.
 \label{eq:band}
\end{equation}
Every member is entangled, $\Phi$ belongs to the family, and every $\omega_f$ in the interval has positive coefficients of all three $\Phi_\alpha$. Thus this infinite family is complete. The following proposition specifies intervals on which one fixed protocol works for all independently chosen ordered inputs.

\begin{proposition}
\label{prop:infinite}
Let $1/2<f_-<f_+<1$ and $f_+<G(f_-)$, with $G$ defined in Eq.~\eqref{eq:recurrence}. The 3-to-1 LOCC protocol in the proof of Proposition~\ref{prop:counter} is universal with ordered fidelity on $\mathcal T_{f_-,f_+}$, succeeds with probability at least $1/2$ on every allowed input, and strictly increases the highest input fidelity whenever $\rho_1\ne\Phi$.
\end{proposition}
Such intervals exist because Eq.~\eqref{eq:strictgain} implies $G(f_-)>f_-$ for $1/2<f_-<1$.

\begin{proof}
We show that $f_+<G(f_-)$ makes the recurrence output from pairs~2 and~3 have greater fidelity than every allowed first-pair state $\omega_z$, where $f_-\le z\le f_+$.

To make this comparison, we first need the recurrence output for unequal fidelities $x$ and $y$. Applying the same map $\mathcal R$ from Eq.~\eqref{eq:recurrencemap} and the matrix-element calculation in Eq.~\eqref{eq:Rentries} gives the unnormalized output on the retained pair,
\begin{align}
 R_{x,y}&:=\mathcal R(\omega_x\otimes\omega_y)\nonumber\\
 &=\begin{pmatrix}
 2a_xa_y&0&0&2c_xc_y\\
 0&2e_xe_y&0&0\\
 0&0&2e_xe_y&0\\
 2c_xc_y&0&0&2a_xa_y
 \end{pmatrix},
 \label{eq:unequalR}
\end{align}
in the pair basis $\{\ket{00},\ket{01},\ket{10},\ket{11}\}$. For $u\in\{x,y\}$, the entries use
\begin{equation}
 e_u:=\frac{1-u}{3},\qquad
 a_u:=\frac{u+e_u}{2},\qquad
 c_u:=\frac{u-e_u}{2}.
 \label{eq:unequalentries}
\end{equation}
As in the equal-fidelity case, the trace gives the success probability of the recurrence operation itself:
\begin{align}
 D(x,y)&:=\tr R_{x,y}=4(a_xa_y+e_xe_y)\nonumber\\
 &=(x+e_x)(y+e_y)+4e_xe_y.
 \label{eq:unequalD}
\end{align}
Its unnormalized overlap with $\Phi$, needed to calculate the conditional fidelity, is
\begin{align}
 N(x,y)&:=\tr(\Phi R_{x,y})
 =2(a_xa_y+c_xc_y)\nonumber\\
 &=xy+e_xe_y.
 \label{eq:unequalN}
\end{align}
Dividing by the success probability gives the fidelity of the normalized recurrence output:
\begin{equation}
 G(x,y):=F\!\left(\frac{R_{x,y}}{D(x,y)}\right)
 =\frac{10xy-x-y+1}{8xy-2x-2y+5}.
 \label{eq:unequal}
\end{equation}
Here $G(x,y)=N(x,y)/D(x,y)$ is the standard recurrence fidelity for unequal isotropic inputs~\cite{Bennett1996,ZangNetworks2025}. On the diagonal, these definitions reduce to those used above: $D(f,f)=D(f)$, $N(f,f)=N(f)$, and $G(f,f)=G(f)$.

We require a lower bound on $G(x,y)$ that holds for all independent choices $x,y\in[f_-,f_+]$. Differentiation gives, for $x,y\in[1/2,1]$,
\begin{equation}
 \partial_xG(x,y)=
 \frac{3(-4y^2+14y-1)}{(8xy-2x-2y+5)^2}>0.
 \label{eq:derivative}
\end{equation}
The expression $8xy-2x-2y+5$ is at least $5$ on this square, so the denominator is strictly positive. The quadratic in the numerator is concave and has values $5$ and $9$ at $y=1/2$ and $y=1$, respectively, so it is positive throughout the interval. Since $G$ is symmetric in its two arguments, it is increasing in each. In particular, the smallest recurrence fidelity on $[f_-,f_+]^2$ is $G(f_-,f_-)=G(f_-)$.

We can now verify the same protocol on every ordered three-pair input from $\mathcal T_{f_-,f_+}$. There are two cases, determined by the state of the first pair.

\emph{Case 1: $\rho_1=\Phi$.} The measurements in step~2 always give $u_1=v_1$, so the branch $\xi=1$ always fails. The successful output is again $\E(\Phi\otimes\rho_2\otimes\rho_3)=\Phi/2$, independently of the allowed states of pairs~2 and~3. Thus the success probability is $1/2$ and the conditional fidelity is one.

\emph{Case 2: $\rho_1=\omega_z$ with $z\in[f_-,f_+]$.} Ordering excludes $\Phi$ from pairs~2 and~3. The joint input is therefore $X_{z;x,y}:=\omega_z\otimes\omega_x\otimes\omega_y$ with $f_-\le y\le x\le z\le f_+$. The monotonicity just established and the choice of the interval give
\begin{equation}
 G(x,y)\ge G(f_-,f_-)=G(f_-)>f_+\ge z.
 \label{eq:bandgain}
\end{equation}
Thus the recurrence output exceeds the fidelity of the first pair for every allowed choice of the states of pairs~2 and~3.

Conditional on $\xi=1$, step~2 gives unequal outcomes with probability $q(z)=2(1-z)/3>0$ and leaves the remaining input $\omega_x\otimes\omega_y$ unchanged. The two successful branches consequently contribute
\begin{equation}
 \E(X_{z;x,y})
 =\tfrac12\omega_z+\tfrac12q(z)R_{x,y}.
 \label{eq:bandoutput}
\end{equation}
The branch retaining pair~1 again contributes zero to the fidelity difference. For the other branch, Eq.~\eqref{eq:bandgain} yields
\begin{align}
 \Delta_{\E}(X_{z;x,y};z)
 &=\frac{q(z)}2[N(x,y)-zD(x,y)]\nonumber\\
 &=\frac{q(z)D(x,y)}2[G(x,y)-z]>0.
 \label{eq:bandgap}
\end{align}
This is the strict form of the universality condition on every input in Case~2. The total success probability is the trace of the two-branch output in Eq.~\eqref{eq:bandoutput}:
\begin{align}
 p_{\rm succ}(z;x,y)
 &:=p_{\E}(X_{z;x,y})\nonumber\\
 &=\frac{1+q(z)D(x,y)}2.
 \label{eq:bandprobability}
\end{align}
The fidelity with $\Phi$ conditioned on the combined success event is
\begin{align}
 F_{\rm out}(z;x,y)
 &:=F_{\E}(X_{z;x,y})\nonumber\\
 &=\frac{z+q(z)N(x,y)}{1+q(z)D(x,y)}.
 \label{eq:bandfidelity}
\end{align}
The gain over the first-pair fidelity $z$ is therefore
\begin{align}
 F_{\rm out}(z;x,y)-z
 &=\frac{q(z)D(x,y)[G(x,y)-z]}
        {1+q(z)D(x,y)}>0.
 \label{eq:bandtotal}
\end{align}
The two cases establish universality on $\mathcal T_{f_-,f_+}$, with equality when $\rho_1=\Phi$ and a strict increase otherwise. The choice $\xi=0$ ensures success probability at least $1/2$ on every input, completing the proof.
\end{proof}

To illustrate Proposition~\ref{prop:infinite} with an explicit interval, choose $f_-=3/4$ and $f_+=25/32$, since $G(3/4)=41/52>25/32$. The resulting family permits the three fidelities to vary independently within $[3/4,25/32]$, subject to ordering, as well as allowing $\Phi$ in any position consistent with that ordering. This gives a concrete infinite complete family admitting ordered universal LOCC purification.

The sufficient condition in Proposition~\ref{prop:infinite} requires $f_+-f_-<G(f_-)-f_-$. By Eq.~\eqref{eq:recurrence}, $G(f)-f\sim(1-f)/3$ as $f\to1^-$, so the permitted interval narrows as its lower bound approaches unity. This restriction concerns the sufficient condition for the present construction and does not establish a necessary bound for other LOCC protocols.

The finite and infinite examples are compatible with Theorem~\ref{thm:main} because neither family supplies states other than $\Phi$ with fidelity approaching one. They consequently omit the approach to $\Phi$ required by the theorem and do not satisfy the segment condition. Their existence establishes that completeness alone is insufficient for the general LOCC no-go conclusion. It does not make convexity a necessary hypothesis for impossibility: the segment condition can also hold in nonconvex families, and Theorem~\ref{thm:main} itself only requires the specified sequence approaching $\Phi$ with a fixed full-rank state on the remaining systems.
\section{Conclusions and Discussions}
\label{sec:discussion}

We have shown that fidelity ordering cannot enable nontrivial universal entanglement purification when independently prepared qubit pairs otherwise range over the full family of states above a fixed lower fidelity threshold below unity. For every fixed finite number of pairs, a separable success map that never lowers the highest input fidelity must preserve that fidelity whenever it succeeds on an allowed input. The same impossibility holds with a fixed finite-dimensional pure catalyst returned exactly upon success. In contrast, an explicit 3-to-1 LOCC protocol achieves nontrivial universal purification on complete finite and infinite families with restricted fidelity ranges. The unordered corollary resolves the higher-threshold universality question for the full two-qubit threshold family, and the ordered result shows that fidelity ordering alone does not permit a strict improvement. The positive constructions establish that completeness by itself cannot determine whether general LOCC permits universal purification.

The no-go theorem gives a useful criterion for deciding what information can justify a purification procedure. Retaining the pair of highest known fidelity already satisfies the universal nondecrease requirement with unit success probability. When the only state constraints are the ordering and a lower fidelity bound below unity, processing additional pairs cannot improve on this conditional fidelity while retaining the same requirement on every allowed input. Within the same finite-input LOCC setting, a proposal for universal improvement must therefore identify additional restrictions on the source or change the performance requirement. This does not mean that complete state tomography is necessary. The constructive examples show that the isotropic form and the specified fidelity ranges in Sec.~\ref{sec:counterexample} can suffice, even when the individual fidelities vary independently. The distinction guides the choice of source information to establish before a purification protocol is applied.

The impossibility follows from the restriction on separable success maps in Theorem~\ref{thm:main}. Fidelity nondecrease on a sequence $\rho_{t_m}\otimes\tau$, with $t_m\to1$ and one fixed full-rank $\tau$, forces every product Kraus operator to have the form in Eq.~\eqref{eq:rigidkraus} on the whole input space. Consequently, the operation preserves the first pair's fidelity for arbitrary independent states of that pair and the remaining systems whenever success has positive probability. Neither success at $\Phi\otimes\tau$ nor the inclusion of $\Phi$ in the allowed family is required as a hypothesis. A fixed Bell-diagonal direction with positive coefficients for all three orthogonal Bell states supplies the sequence used in the proof. This explains why arbitrarily small success probabilities and exclusion of $\Phi$ from the allowed family do not remove the restriction.

The universal requirement differs from limits on purifying a specified resource state. General resource-purification theorems bound the error in converting a noisy input into a pure resource, including under probabilistic operations, and extend these bounds to channels~\cite{FangLiu2020,FangLiu2022}. Here we require one fixed map to avoid a fidelity loss on every allowed joint input formed from independently prepared pairs, and infer equality with the highest input fidelity. Quantification over the input family is essential. For a specified entangled two-qubit state, asymptotic purification is possible when sufficiently many identical copies are available~\cite{Horodecki1997Distillation}. Universal concentration under an identical-pure-state assumption and purification of specified correlated resources also concern different domains~\cite{MatsumotoHayashi2007,BuscemiDatta2010}. Our requirement is distinct from optimizing a network rate~\cite{Goodenough2021,LiCutoffs2021}, and the target Bell state is fixed and known rather than an unknown pure state~\cite{Zhao2026}. Bounds on the asymptotic purification rate~\cite{WangDuan2016} and state-specific optimization of teleportation fidelity~\cite{VerstraeteVerschelde2003} address performance criteria distinct from the exact family-wide inequality imposed here.

For LOCC or separable operations with catalytic assistance, Lemma~\ref{lem:catalyst} gives a separate reduction: exact return of a finite-dimensional pure catalyst on one full-rank input forces the induced success map to have a product Kraus representation. The lemma applies to the separable, trace-nonincreasing map selected on success and imposes no return condition after failure. Earlier work already excludes catalytic fidelity improvement for certain single-copy mixture families under a protocol independent of the mixture parameter~\cite{EisertWilkens2000}. The present reduction uses exact pure-state return on one full-rank input to establish separability of the induced map, which can then be combined with the ordered $n$-to-1 theorem. The characterization of noise-robust catalytic channels uses trace-preserving processes with return on a neighborhood of inputs~\cite{Son2026}. Purity and full rank play distinct roles here: purity makes the returned catalyst uncorrelated with the successful output, and full rank extends this factorization to all input vectors. The Schmidt-rank argument then yields separability of the induced map. This result is compatible with catalytic transformations of rank-deficient states~\cite{JonathanPlenio1999,EisertWilkens2000}. Conditions on Schmidt coefficients characterize exact catalytic conversions between specified pure bipartite states~\cite{DaftuarKlimesh2001,Turgut2007}. Such conversions need not return a catalyst on any full-rank input, as required by Lemma~\ref{lem:catalyst}.

Other results on catalytic purification address different assumptions. PPT bound entangled states cannot yield Bell pairs at a positive asymptotic rate, even with correlated catalysts~\cite{LamiRegulaStreltsov2024}, whereas one-shot resource purification can benefit from catalysts when the allowed overhead, accuracy, and catalytic model are changed~\cite{FangLiu2026}. In that one-shot setting, the protocol and catalyst are chosen for a specified source-state conversion. The catalyst may be mixed and may remain correlated with the successful output even though its reduced state is returned exactly. Our theorem instead requires a fixed pure catalyst and a fixed operation to serve the entire ordered input family. Allowing correlations with a catalyst while approximating a desired output leads to different conditions for pure-state transformations~\cite{KondraDattaStreltsov2021}, and catalytic assistance can improve teleportation performance for specified resource states~\cite{LipkaBartosikSkrzypczyk2021}. Approximate return also permits entanglement embezzlement, in which entanglement is extracted while the catalyst changes by an arbitrarily small amount as its dimension grows~\cite{VanDamHayden2003,ZanoniTheurerGour2024}. These results illustrate why pure, mixed, correlated, and approximately returned catalysts must be distinguished~\cite{LipkaBartosik2024}.

Existing guarantees for nonidentical pairs depend on both the allowed family and the quantity used for comparison. The recurrence protocol is universal on the incomplete family described at the start of Sec.~\ref{sec:counterexample}~\cite{ZangNetworks2025}. For isotropic inputs, the same work gives guarantees relative to an average or a specified weighted combination of the input fidelities. These bounds do not require the output to attain the higher input fidelity for every allowed pair. Accumulating and shuffling pairs from heterogeneous sources can also improve the expected success probability and success-weighted output fidelity relative to processing without shuffling~\cite{ZangThesis2026}. This comparison uses processing without shuffling as its baseline, rather than the highest input fidelity.

The positive constructions in Propositions~\ref{prop:counter} and~\ref{prop:infinite} establish ordered universality on complete families containing both $\Phi$ and full-rank states. The contribution of these constructions lies in the three-pair success and output rules combined with the specified input families. The bilateral-CNOT recurrence operation and its fidelity formulas are standard~\cite{Bennett1996,ZangNetworks2025}. The constructed protocol retains the first pair on one branch. On the other branch, it requires unequal computational-basis outcomes on that pair before applying recurrence to the remaining two pairs. When the first pair is $\Phi$, this second branch has zero success probability, so the successful output retains fidelity one. For the other allowed inputs, the restrictions on the fidelities ensure that the recurrence output has fidelity greater than that of the first pair. The combined success operation therefore never lowers the highest input fidelity and sometimes increases it.

These complete families have all nonunit fidelities bounded away from one and hence omit the sequence required by Theorem~\ref{thm:main}. Their existence establishes that the measurement and output rules of the specified biCEP class are essential to its no-go result~\cite{Zang2025}. Clifford gates alone do not impose that result on general LOCC protocols. Completeness together with convexity is sufficient to obtain the segment condition used here, but the examples do not make convexity necessary for impossibility. The constructions assume ideal local operations and establish existence without claiming optimal success probability, fidelity gain, or network rate. Optimizing these quantities within the permitted families is a separate problem for which existing purification-optimization methods provide relevant tools~\cite{Rozpedek2018,Krastanov2019}.

A remaining problem is to characterize the input families that admit a strict universal improvement. The sufficient conditions for impossibility and the explicit possible families established here do not provide a necessary-and-sufficient criterion. A related practical question is the tradeoff among the maximal fidelity loss permitted on any allowed input, the improvement on favorable inputs, and the success probability. Our theorem determines the exact zero-loss case for the full threshold family, but does not bound these quantities when a nonzero loss is permitted. Higher-dimensional targets, correlations between input pairs, mixed catalysts, and approximate catalyst return also require additional analysis. These directions use the no-go theorem as a precise reference point for assessing which changes in source assumptions or performance requirements can support useful finite-input purification.

The role of source information also suggests questions in other quantum resource theories~\cite{ChitambarGour2019}. In coherence theory, the free states are diagonal in a specified orthonormal reference basis~\cite{Streltsov2017}. Necessary-and-sufficient conditions characterize deterministic conversion of a specified mixed coherent state into a pure state under strictly incoherent operations~\cite{LiuZhou2019}. These conditions concern a given resource, whereas a universal protocol must satisfy its performance requirement throughout a prescribed family without depending on its individual members.

For a concrete extension, fix the qubit reference basis $\{\ket0,\ket1\}$ and define $\ket+:=(\ket0+\ket1)/\sqrt2$ and $F_{\mathrm{coh}}(\rho):=\bra+\rho\ket+$ for a single-qubit state $\rho$. The comparison concerns a fidelity with a specified maximally coherent state rather than a coherence monotone. One may ask whether a fixed $n$-to-1 protocol can preserve or increase the highest input value of $F_{\mathrm{coh}}$ whenever it succeeds, with strict improvement for some allowed input, for independently prepared qubits with a known ordering of these fidelities. A precise operational choice is a success map with strictly incoherent Kraus matrices, each having at most one nonzero entry in every row and column in the tensor-product reference basis of the input qubits and the reference basis of the output qubit~\cite{Streltsov2017,LiuZhou2019}. The product Kraus classification proved here relies on bipartite separability and does not establish this coherence analogue. Determining its answer requires a separate analysis of the allowed coherence operations and input families.

\begin{acknowledgments}
This work was supported by the National Natural Science Foundation of China (Grant No.~12405019) and the Natural Science Foundation of Shandong Province (Grant No.~ZR2025LLZ002).
\end{acknowledgments}

\let\originalbibfont\bibfont
\renewcommand{\bibfont}{\originalbibfont\interlinepenalty=10000\relax}

\end{document}